\documentclass{article}

\PassOptionsToPackage{numbers, compress}{natbib}

\usepackage[final]{neurips_2023}
\makeatletter
\def\@noticestring{}
\makeatother

\usepackage[utf8]{inputenc}
\usepackage[T1]{fontenc}
\usepackage{xurl}
\usepackage{hyperref}
\usepackage{booktabs}
\usepackage{amsfonts}
\usepackage{nicefrac}
\usepackage{microtype}
\usepackage{xcolor}
\usepackage{graphicx}
\usepackage{tabularx}

\title{Towards best practices in AGI safety \\ and governance: A survey of expert opinion}

\author{
Jonas Schuett\thanks{Contact: jonas.schuett@governance.ai} \quad Noemi Dreksler \quad Markus Anderljung \quad David McCaffary \\\\
\quad \textbf{Lennart Heim} \quad \textbf{Emma Bluemke} \quad \textbf{Ben Garfinkel} \\
\\ Centre for the Governance of AI
}

\begin{document}

\maketitle

\begin{abstract}
A number of leading AI companies, including OpenAI, Google DeepMind, and Anthropic, have the stated goal of building artificial general intelligence (AGI)—AI systems that achieve or exceed human performance across a wide range of cognitive tasks. In pursuing this goal, they may develop and deploy AI systems that pose particularly significant risks. While they have already taken some measures to mitigate these risks, best practices have not yet emerged. To support the identification of best practices, we sent a survey to 92 leading experts from AGI labs, academia, and civil society and received 51 responses. Participants were asked how much they agreed with 50 statements about what AGI labs should do. Our main finding is that participants, on average, agreed with all of them. Many statements received extremely high levels of agreement. For example, 98\% of respondents somewhat or strongly agreed that AGI labs should conduct pre-deployment risk assessments, dangerous capabilities evaluations, third-party model audits, safety restrictions on model usage, and red teaming. Ultimately, our list of statements may serve as a helpful foundation for efforts to develop best practices, standards, and regulations for AGI labs.

\newpage
\paragraph{Key findings}

\begin{itemize}
  \item There was a broad consensus that AGI labs should implement most of the safety and governance practices in a 50-point list. For every practice but one, the majority of respondents somewhat or strongly agreed that it should be implemented. Furthermore, for the average practice on our list, 85.2\% somewhat or strongly agreed it should be implemented.
  \item Respondents agreed especially strongly that AGI labs should conduct pre-deployment risk assessments, dangerous capabilities evaluations, third-party model audits, safety restrictions on model usage, and red teaming. 98\% of respondents somewhat or strongly agreed that these practices should be implemented. On a numerical scale, ranging from -2 to 2, each of these practices also received a mean agreement score of at least 1.76.
  \item Experts from AGI labs had higher average agreement with statements than respondents from academia or civil society. However, no significant item-level differences were found.
\end{itemize}

\paragraph{Policy implications}

\begin{itemize}
  \item AGI labs can use our findings to conduct an internal gap analysis to identify potential best practices that they have not yet implemented. For example, our findings can be seen as an encouragement to make or follow through on commitments to commission third-party model audits, evaluate models for dangerous capabilities, and improve their risk management practices.
  \item In the US, where the White House has recently expressed concerns about the dangers of AI, regulators and legislators can use our findings to prioritize different policy interventions. In the EU, our findings can inform the debate on to what extent the proposed AI Act should account for general-purpose AI systems. In the UK, our findings can be used to draft upcoming AI regulations as announced in the recent White Paper “A pro-innovation approach to AI regulation”.
  \item Our findings can inform an ongoing initiative of the Partnership on AI to develop shared protocols for the safety of large-scale AI models. They can also support efforts to adapt the NIST AI Risk Management Framework and ISO/IEC 23894 to developers of general-purpose AI systems. Finally, they can inform the work of CEN-CENELEC to develop harmonized standards for the proposed EU AI Act, especially on risk management.
  \item Since most practices are not inherently about AGI labs, our findings might also be relevant for other organizations that develop and deploy increasingly general-purpose AI systems, even if they do not have the goal of building AGI.
\end{itemize}

\end{abstract}

\newpage
\setcounter{footnote}{0}

\section{Introduction}
\label{section:Introduction}

\paragraph{Background.} Over the past few months, a number of powerful artificial intelligence (AI) systems were released \cite{openai:2023a, pichai:2023, touvron:2023} and integrated into products that are now being used by millions of people around the world \cite{mehdi:2023, spataro:2023, wright:2023}. At the same time, some leading AI companies have become more explicit that their ultimate goal is to build artificial general intelligence (AGI)—AI systems that achieve or exceed human performance across a wide range of cognitive tasks \cite{altman:2023, kruppa:2023, perrigo:2023}. The prospect of AGI used to be a fringe area \cite{goertzel:2007, goertzel:2014, bostrom:2014}, but the debate has now entered the public discourse \cite{yudkowsky:2023, hogarth:2023, klein:2023, metz:2023} and the political stage \cite{whitehouse:2023b, ukdepartment:2023, bertuzzi:2023a, solender:2023}.\footnote{In some cases, policymakers use the term “AGI” explicitly \cite{hmgovernment:2021, ukdepartment:2023}. In other cases, they talk about developers of “general-purpose AI systems” and “foundation models” \cite{bertuzzi:2023a, bertuzzi:2023b} or “generative AI systems” \cite{whitehouse:2023b, solender:2023}, which also includes organizations that try to build AGI.} There are now increasing efforts to develop standards and regulations that would apply to organizations that try to build AGI. However, there are still a number of open questions about the substance of such standards and regulations.

\paragraph{Purpose.} This paper is intended to contribute to the creation of best practices in AGI safety and governance. We want to make sure that the views of relevant experts are taken into account. More specifically, we want to find out which practices already have broad support and where more work is needed. To this end, we surveyed 51 leading experts from AGI labs, academia, and civil society. Our findings can be used as evidence in discussions about the creation of best practices. We hope that AGI labs will follow emerging best practices on a voluntary basis. But best practices could also inform standard-setting processes (e.g. by ISO and NIST) and regulatory efforts. Consider the following simple model of how governance mechanisms get codified into law: (1) different companies experiment with different governance mechanisms; (2) best practices emerge; (3) best practices inform standard-setting processes; (4) standards get codified into law. The main purpose of this paper is to support step (2). However, in practice, these steps are often performed in parallel, not in a sequential way. The paper could therefore also inform steps (3) and (4).

\paragraph{Related work.} AGI labs share some information about their governance practices \cite{brundage:2022, ganguli:2022, kavukcuoglu:2022, openai:2023a} and occasionally even propose best practices \cite{cohere:2022}. There do not seem to be any independent efforts to create best practices for the governance of organizations that try to build “AGI”. However, there are efforts that target developers of “general-purpose AI systems”, “foundation models”, or “large-scale AI models”, which also includes AGI labs. Most notably, the Partnership on AI has initiated a multistakeholder dialogue to develop shared protocols for the safety of large-scale AI models \cite{partnershiponai:2023}, while The Future Society seeks to create an industry code of conduct for developers of general-purpose AI systems and foundation models \cite{thefuturesociety:manuscript}. There are also efforts to adapt AI risk management standards like the NIST AI Risk Management Framework \cite{nist:2023} or ISO/IEC 23894 \cite{iso23894:2023} to the needs of developers of general-purpose AI systems \cite{barrett:nd}. The Alignment Research Center (ARC) is also developing a new standard on dangerous capabilities evaluations that is targeted at “leading AI companies” \cite{arc:2023}. Finally, the proposed EU AI Act will likely contain rules for developers of general-purpose AI systems and foundation models \cite{bertuzzi:2023b}, though the issue remains disputed \cite{ainowinstitute:2023}.

\paragraph{Terminology.} By “AGI”, we mean AI systems that reach or exceed human performance across a wide range of cognitive tasks.\footnote{There is no generally accepted definition of the term “AGI”. According to Goertzel \cite{goertzel:2011}, the term was first used by Gubrud \cite{gubrud:1997} in the article “Nanotechnology and international security”. It was popularized through the book “Artificial general intelligence” edited by Goertzel and Pennachin \cite{goertzel:2007}. We acknowledge that our definition is vague. For more information on how to make this definition more concrete, we refer to the relevant literature \cite{goertzel:2014, muehlhauser:2013, barnett:2020}. Different definitions emphasize different elements. For example, in their charter, OpenAI uses a definition that focuses on economic value: “highly autonomous systems that outperform humans at most economically valuable work” \cite{openai:2018}. But note that they have recently used a simplified definition: “AI systems that are generally smarter than humans” \cite{altman:2023}. The term “AGI” is related to the terms “strong AI” \cite{searle:1980}, “superintelligence” \cite{bostrom:1998, bostrom:2014}, and “transformative AI” \cite{gruetzemacher:2019}.}
(Note that we do not make any claims about when, if at all, AGI will be built.)\footnote{For an overview of different methods to forecast AI progress, see \cite{wynroe:2023}.} By “AGI labs”, we mean organizations that have the stated goal of building AGI. This includes OpenAI, Google DeepMind, and Anthropic. Since other AI companies like Microsoft and Meta conduct similar research (e.g. training very large models), we also refer to them as “AGI labs” in this paper. By “AGI safety and governance practices”, we mean internal policies, processes, and organizational structures at AGI labs intended to reduce risk.

\paragraph{Overview.} The paper proceeds as follows. Section~\ref{section:Methods} contains information about the sample, the survey, and our analysis. Section~\ref{section:Results} reports our results, namely to what extent respondents agreed with different statements about what AGI labs should do, whether there were noticeable differences between sectors and genders, and which additional practices respondents suggested. Figure \ref{fig:figure_1} shows the percentages of responses for all statements listed in the survey. Section~\ref{section:Discussion} discusses our key results, their policy implications, and the main limitations of our study. It also suggests directions for future work. Section~\ref{section:Conclusion} concludes. Appendix~\ref{section:A} contains a list of all participants who gave us permission to mention their names and affiliations. Appendix~\ref{section:B} contains a list of all statements used in the survey. Appendices~\ref{section:D}, ~\ref{section:E}, and ~\ref{section:F} contain additional figures, tables, and analyses.

\section{Methods}
\label{section:Methods}

\subsection{Sample}
\label{section:Sample}

\paragraph{Sample size.} We invited 92 experts to take the survey and received 51 responses. The response rate was 55.4\%, which is high compared to previous expert surveys of AI researchers \cite{grace:2018, zhang:2021, steinperlman:2022}.

\begin{figure}[b]
    \centering
    \includegraphics[width=0.9\textwidth]{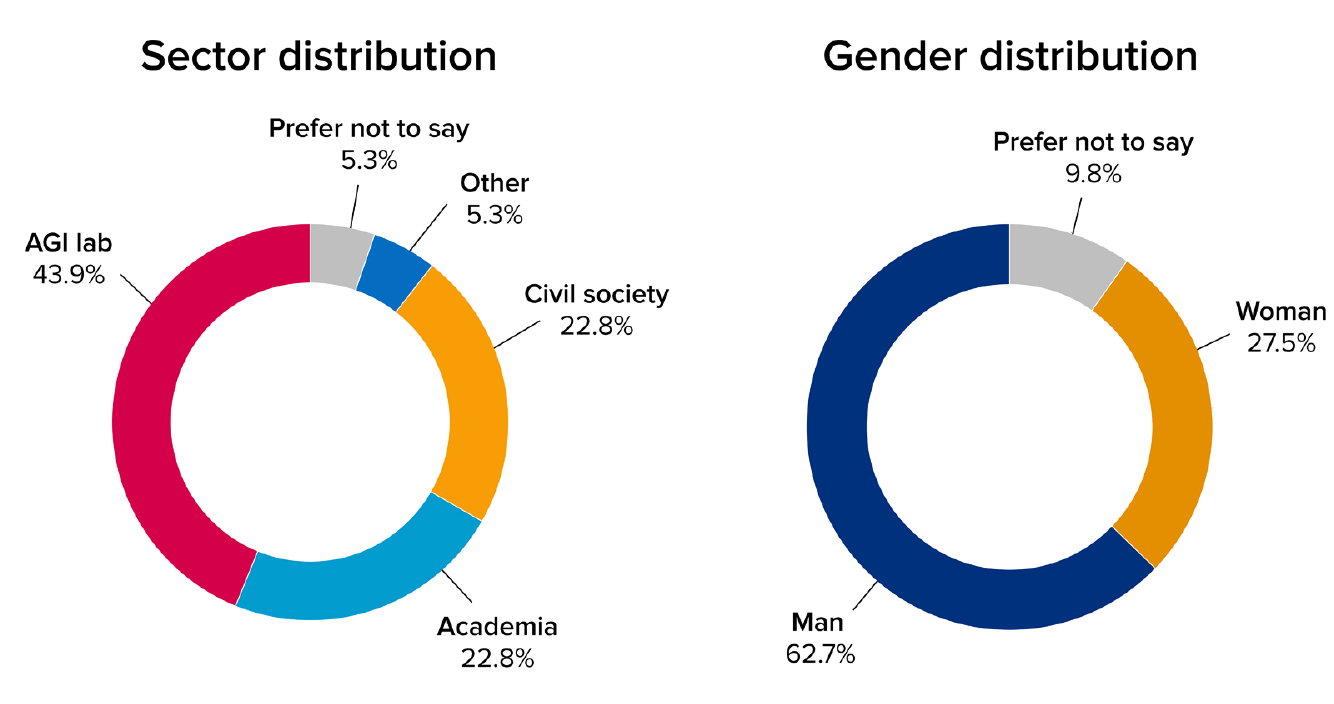}
    \caption{\textbf{Sample by sector and gender} | The figure shows the sector of work and gender of the respondents. Respondents could choose more than one sector in which they work.}
    \label{fig:figure_2}
\end{figure}

\paragraph{Sample selection.} Participants were selected in a four-step approach. In the first step, we selected relevant sectors: AGI labs, academia, civil society (including nonprofit organizations and think tanks), and other (including government, consulting firms, and other tech companies). In the second step, we selected specific organizations within each sector. In the third step, we selected experts within each organization. In the fourth step, we added individual experts who were not affiliated with any of the organizations identified in the second step. The final sample represented all of the selected sectors identified in the first step. Figure \ref{fig:figure_2} shows the division of respondents by sector and gender. 33 respondents (64.7\%) gave us permission to list them publicly as respondents to the survey. The full list can be found in Appendix~\ref{section:A}.

\paragraph{Sample type.} Our sample could best be described as a purposive sample \cite{palinkas:2015}. We selected individual experts based on their knowledge and experience in areas relevant for AGI safety and governance, but we also considered their availability and willingness to participate. We used a number of proxies for expertise, such as the number, quality, and relevance of their publications as well as their role at relevant organizations.

Overall, we believe the selection reflects an authoritative sample of current AGI safety and governance-specific expertise. For a discussion of limitations related to our sample, see Section~\ref{section:Limitations}.

\begin{figure}[!htbp]
    \centering
    \vspace*{-1.6cm}
    \hspace*{-2.5cm}
    \includegraphics[width=1.3\textwidth]{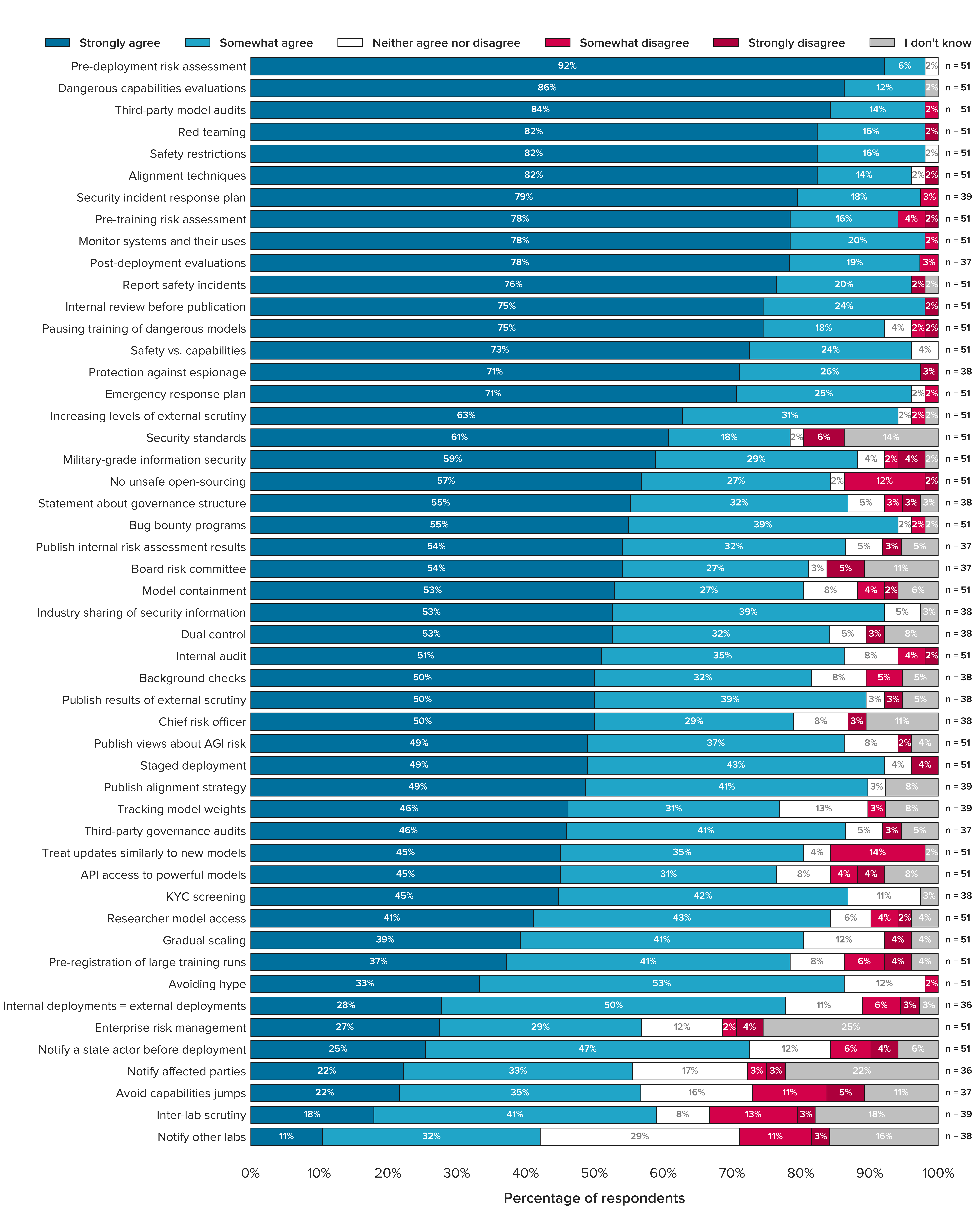}
    \caption{\textbf{Percentages of responses for all statements} | The figure shows the percentage of respondents choosing each answer option. At the end of each bar we show the number of people who answered each item. The items are ordered by the total number of respondents that “strongly” agreed. The full statements can be found in Appendix~\ref{section:B}.}
    \label{fig:figure_1}
\end{figure}

\begin{figure}[!htbp]
    \centering
    \vspace*{-1.6cm}
    \hspace*{-1.2cm}
    \includegraphics[width=1.2\textwidth]{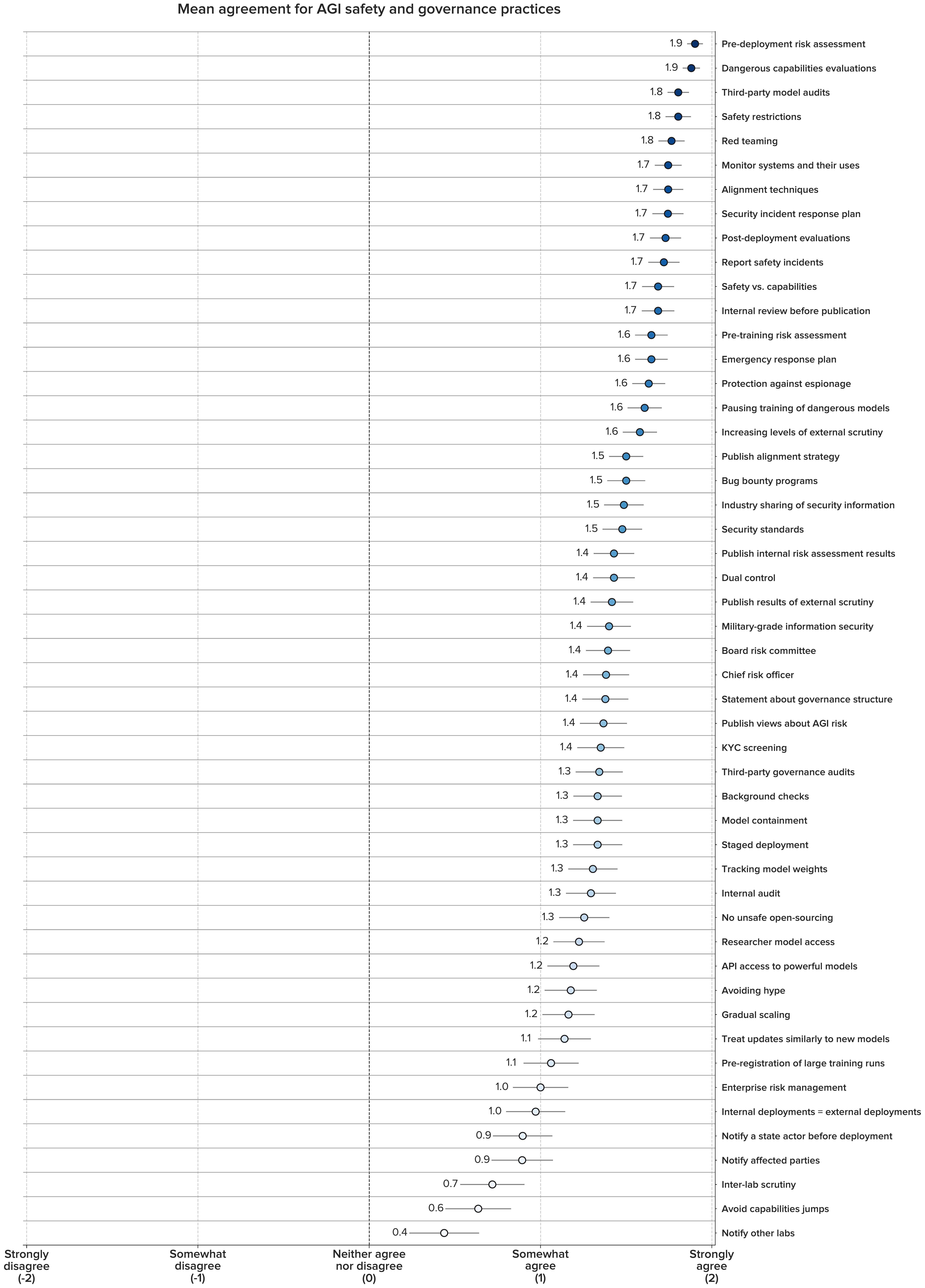}
    \caption{\textbf{Mean agreement for all statements} | The figure shows the mean and 95\% confidence interval for each of the 50 statements. “I don’t know responses” were excluded from the analysis.
}
    \label{fig:figure_4}
\end{figure}

\subsection{Survey}
\label{section:Survey}

\paragraph{Survey design.} Informed consent had to be given before proceeding to the main survey. The survey began by defining the terms “AGI”, “AGI labs”, and “AGI safety and governance practices” as noted above. Respondents were then asked to what extent they agree or disagree with statements about what AGI labs should do. We asked respondents for their gender and where they worked. Finally, respondents were able to list important AGI safety and governance practices they thought were missing from the survey. Respondents took a median of 11 minutes to complete the survey.

\paragraph{Statements about AGI safety and governance practices.} The statements covered many different areas, including development, deployment, monitoring, risk management, external scrutiny, information security, communication, and culture. They were extracted from (1) current practices at individual AGI labs (e.g. pre-deployment risk assessments \cite{brundage:2022, kavukcuoglu:2022} and dangerous capabilities evaluations \cite{openai:2023a}), (2) planned practices at individual labs (e.g. third-party model audits \cite{altman:2023}), (3) proposals in the literature (e.g. third-party governance audits \cite{mokander:2023} and incident reporting \cite{mcgregor:2021}), and (4) discussion with experts and colleagues. In total, the survey contained 50 statements, 30 of which respondents were required to respond to and 20 where answers were optional. Appendix~\ref{section:B} contains a full list of all statements.

\paragraph{Response scale.} Respondents were asked to indicate their level of agreement based on a 5-point Likert scale: “strongly disagree” (-2), “somewhat disagree” (-1), “neither agree nor disagree” (0), “somewhat agree” (1), “strongly agree” (2). They also had the option to say “I don’t know”.

\paragraph{Demographic questions.} Respondents were asked what their gender was (“man”, “woman”, “another gender”, “prefer not to say”) and what sector they worked in (“AGI lab [e.g. OpenAI, Google DeepMind, Anthropic, Microsoft, and Meta]”, “other tech company”, “consulting firm”, “think tank”, “nonprofit organization”, “government”, “academia”, “other”, “prefer not to say”). For the sector question, respondents were able to choose more than one option.

\paragraph{Survey distribution.} The survey took place between 26 April and 8 May 2023. Respondents were sent an initial email invitation and a reminder email using Qualtrics. A one hour virtual workshop was held which invited the same individuals as the sampling frame. The workshop explored questions on how AGI safety and governance practices could be created and implemented. 21 people attended the workshop along with the seven authors of this paper who took notes and moderated the discussion. During the workshop, attendees were reminded to participate in the survey. Additional follow-up emails were sent to respondents in the final three days of the survey in order to ensure the sample was more representative of the sampling frame and that emails had not gone unseen due to email filters that may have flagged the Qualtrics survey invitations and reminder emails as spam.

\paragraph{Anonymity.} Responses to the survey were anonymous. The part of the survey that asked respondents for their views was a separate Qualtrics survey to both the informed consent survey and where respondents noted their name and affiliation. We will not make any of the demographic data or text responses public to further ensure that responses cannot be reverse-identified. Respondents were informed of these measures in the informed consent section.

\subsection{Analysis}
\label{section:Analysis}

\paragraph{Demographic groups.} We categorized sector responses as follows: AGI lab, academia, civil society (“think tank”, “nonprofit organization”), other (“other tech company”, “consulting firm”, “government”, “other”).

\paragraph{Group differences.} To test for differences in the overall population of responses across all items, we used the Mann-Whitney U test. To test for differences between groups in responses for each practice, we used Chi-squared tests. Certain subgroups had to be removed from the gender (“another gender”, “prefer not to say”) and sector (“other”, “prefer not to say”) analyzes due to sample sizes falling below 5 \cite{mcdonald:2009}. Where applicable throughout, the Holm-Bonferroni correction was used to correct for multiple comparisons: the original alpha-value (0.05) is divided by the number of remaining tests, counting down from the highest to the lowest p-value. The p-values were then compared to the Holm-Bonferroni-adjusted significance levels to determine the significance of each test.

\paragraph{Open science.} The survey draft, pre-registration, pre-analysis plan, code, and data can be found on OSF (\url{https://osf.io/s7vhr}). To protect the identity of respondents, we will not make any demographic data or text responses public. We largely followed the pre-analysis plan. Any deviations from the pre-registered analyses can be found in Appendix~\ref{section:F}, along with the pre-registered cluster analysis.

\section{Results}
\label{section:Results}

In this section, we report the main results of the survey, namely respondents’ level of agreement (Section~\ref{section:Level of agreement}), differences between sectors and genders (Section~\ref{section:DifferencesResults}), and additional practices that were suggested by respondents (Section~\ref{section:SuggestedResults}). Additional figures, tables, and analyses can be found in Appendices~\ref{section:D}, ~\ref{section:E}, and ~\ref{section:F}.

\subsection{Level of agreement}
\label{section:Level of agreement}

\paragraph{Overall agreement.} There was a broad consensus that AGI labs should implement most of the safety and governance practices in a 50-point list. For 98\% of the practices, a majority (more than 50\%) of respondents strongly or somewhat agreed. For 56\% of the practices, a majority (more than 50\%) of respondents strongly agreed. The mean agreement across all 50 items was 1.39 on a scale from -2 (strongly disagree) to 2 (strongly agree)—roughly halfway between somewhat agree and strongly agree. On average, across all 50 items, 85.2\% of respondents either somewhat or strongly agreed that AGI labs should follow each of the practices. On average, only 4.6\% either somewhat or strongly disagreed that AGI labs should follow each of the practices. The broad level of agreement can be seen in Figure \ref{fig:figure_1}, which shows the percentage of respondents that answered “strongly agree”, “somewhat agree”, “neither agree nor disagree”, “somewhat disagree”, “strongly disagree”, and “I don’t know” for each of the potential AGI best practices. For none of the practices, a majority (more than 50\%) of respondents somewhat or strongly disagreed. Indeed, the highest total disagreement on any item was 16.2\% for the item “avoid capabilities jumps”. Across all 2,285 ratings respondents made, only 4.5\% were disagreement ratings.

\begin{figure}[!b]
    \centering
    \hspace*{-0cm}
    \includegraphics[width=1\textwidth]{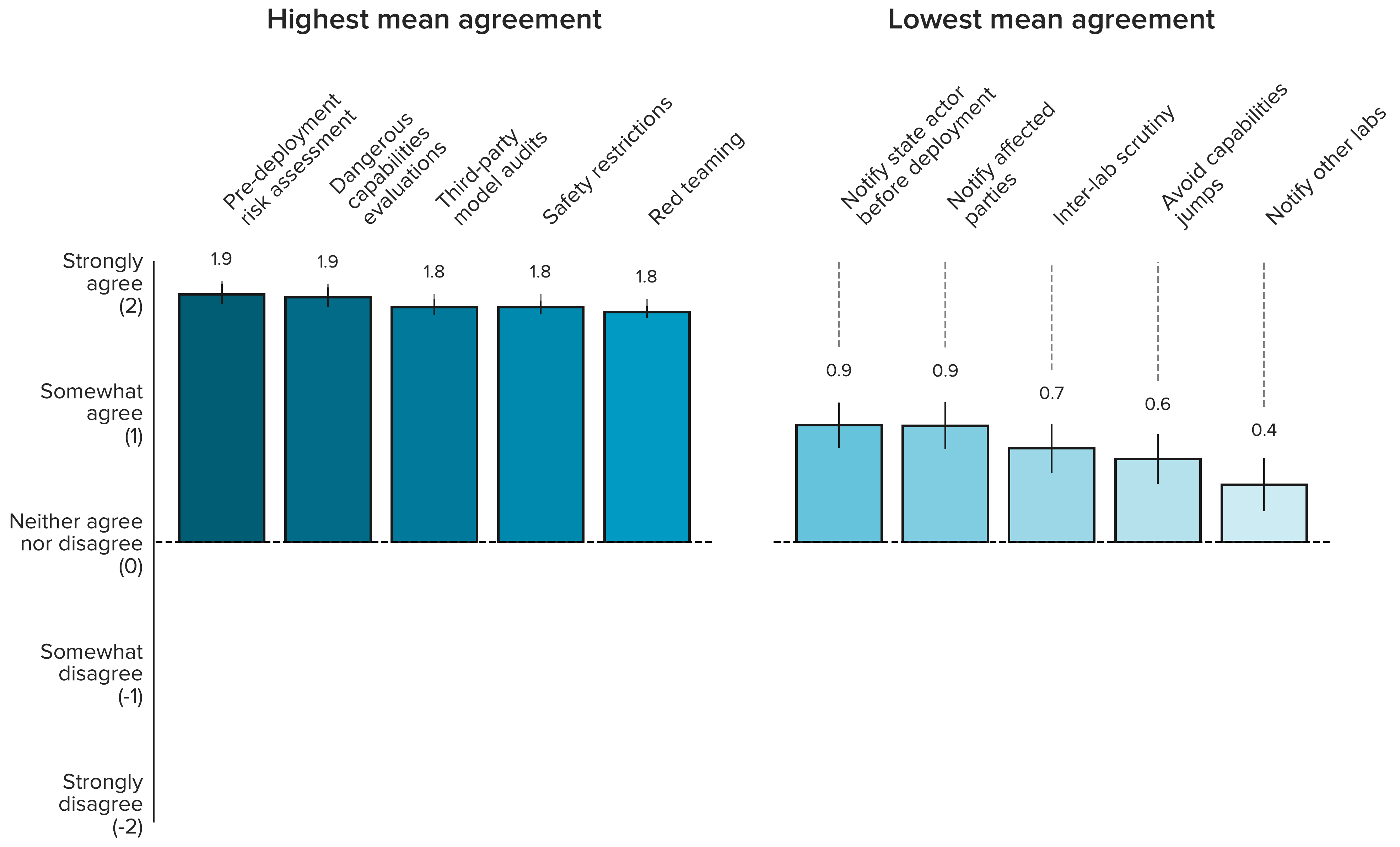}
    \caption{\textbf{Statements with highest and lowest mean agreement} | The figure shows the mean agreement and 95\% confidence interval for the five highest and lowest mean agreement items.}
    \label{fig:figure_3}
\end{figure}

\paragraph{Highest agreement.} The items with the highest total agreement proportions all had agreement ratings from 98\% of respondents were: dangerous capabilities evaluations, internal review before publication, monitor systems and their uses, pre-deployment risk assessment, red teaming, safety restrictions, and third-party model audits. Seven items had no disagreement ratings at all: dangerous capabilities evaluations, industry sharing of security information, KYC screening, pre-deployment risk assessment, publish alignment strategy, safety restrictions, and safety vs. capabilities. Figure \ref{fig:figure_3} shows the statements with the highest and lowest mean agreement. The mean agreement for all statements can be seen in Figure \ref{fig:figure_4}. The statements with the highest mean agreement were: pre-deployment risk assessment (\textit{M} = 1.9), dangerous capabilities assessments (\textit{M} = 1.9), third-party model audits (\textit{M} = 1.8), safety restrictions (\textit{M} = 1.8), and red teaming (\textit{M} = 1.8).

\paragraph{Lowest agreement.} The five items with the highest total disagreement proportions among respondents were: avoid capabilities jumps (16.2\%), inter-lab scrutiny, (15.4\%), no unsafe open-sourcing, (13.7\%), treat updates similarly to new models, (13.7\%), and notify other labs, (13.2\%). The five statements with the lowest mean agreement were: notify other labs (\textit{M} = 0.4), avoid capabilities jumps (\textit{M} = 0.6), inter-lab scrutiny (\textit{M} = 0.7), notify affected parties (\textit{M} = 0.9), and notify a state actor before deployment (\textit{M} = 0.9). Note that all practices, even those with lowest mean agreement, show a positive mean agreement, that is above the midpoint of “neither agree nor disagree” and in the overall agreement part of the scale.

\begin{figure}[!b]
    \centering
    \hspace*{-0cm}
    \includegraphics[width=1\textwidth]{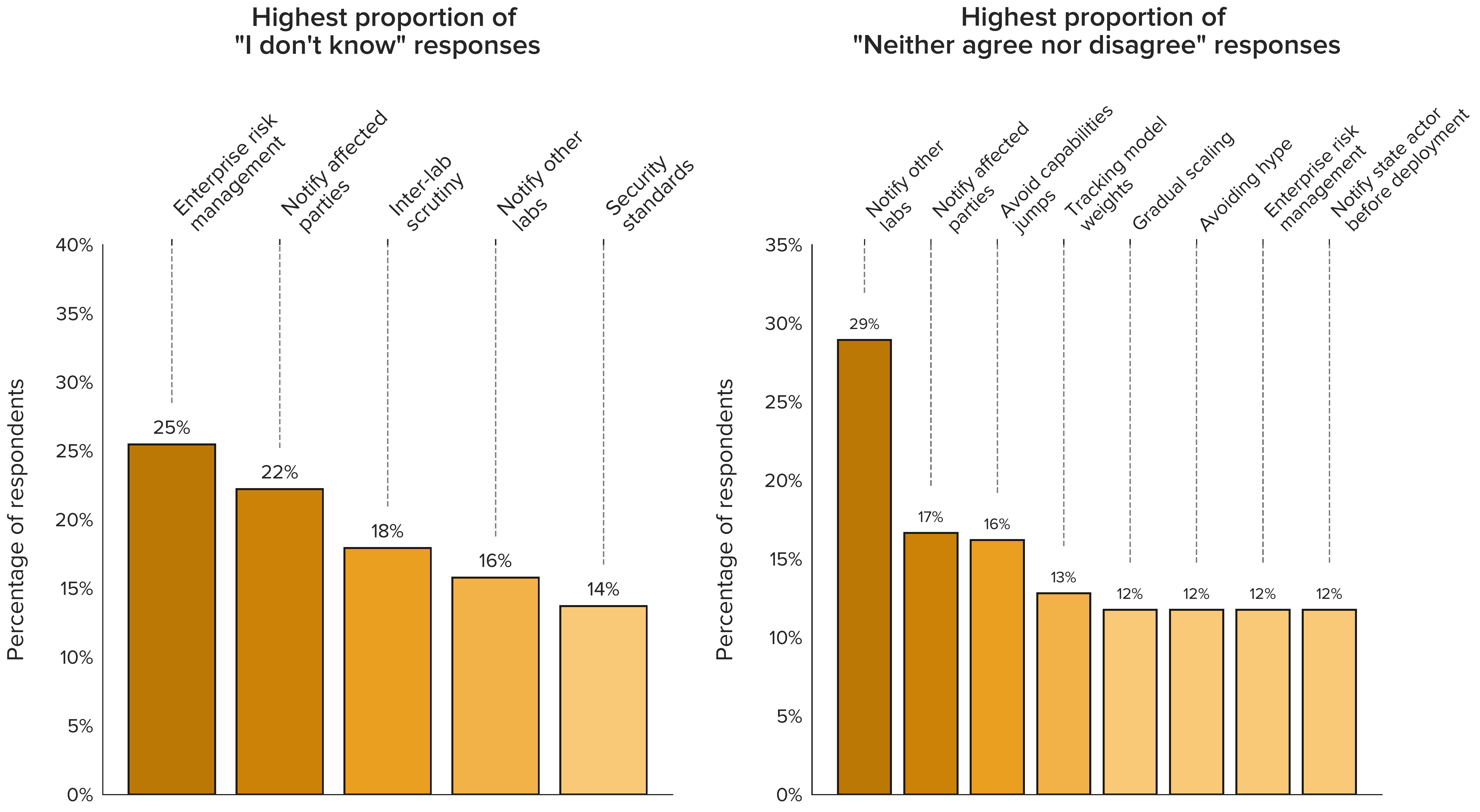}
    \caption{\textbf{Statements with the highest proportion of “I don’t know” and “neither agree nor disagree” responses}}
    \label{fig:figure_5}
\end{figure}

\paragraph{“I don’t know” and “neither agree nor disagree”.} The five practices with the highest proportion of “I don’t know” and “neither agree nor disagree” responses can be seen in Figure \ref{fig:figure_5}. Enterprise risk management (25.5\%), notify affected parties (22.2\%), inter-lab scrutiny (17.9\%), notify other labs (15.8\%), and security standards (13.7\%) show the highest “I don’t know” responses. The four practices with the highest “neither agree nor disagree” responses were: notify other labs (28.9\%), notify affected parties (16.7\%), avoid capabilities jumps (16.2\%), and tracking model weights (12.8\%). Avoiding hype, enterprise risk management, gradual scaling, and notify a state actor before deployment are all tied for fifth highest “neither agree nor disagree” responses (11.8\%).

\subsection{Differences between sectors and genders}
\label{section:DifferencesResults}

\paragraph{Statistical tests.} We used two statistical tests to test for differences between sectors and genders. Firstly, we conducted Mann-Whitney U tests to test for differences in the overall mean agreement across all items. This is a test of whether two independent samples are drawn from the same underlying distribution, and does not assume that this underlying distribution is normal, making it an appropriate test statistic for our data. Secondly, we conducted Chi-squared tests of independence to test for significant differences in the distribution of agreement and disagreement responses for each item by gender and sector. This test compares the observed frequencies across the categories of interest with the frequencies which would be expected if there was no difference between the responses in each category.

\paragraph{Differences between sectors.} We found a significant difference in overall mean agreement across items between respondents from AGI labs and academia (U = 325295.0, p < 0.001, $\alpha $ = 0.017), as well as between respondents from AGI labs and civil society (U = 1106715.0, p < 0.001, $\alpha $ = 0.017). Respondents from AGI labs (\textit{M} = 1.54) showed significantly higher mean agreement than respondents from academia (\textit{M} = 1.16) and civil society (\textit{M} = 1.36). There was no significant difference in overall mean agreement between academia and civil society. When comparing sector groups at the item level we found no significant differences between sector groups for any of the items. The mean agreement by sector can be seen in Figures \ref{fig:figure_6} and \ref{fig:figure_7} in Appendix~\ref{section:D}.

\paragraph{Differences between genders.} We found no significant differences between responses from men and women—neither in overall mean agreement, nor at the item level. The mean agreement by gender can be seen in Figure \ref{fig:figure_8} in Appendix~\ref{section:D}.

\subsection{Suggested practices}
\label{section:SuggestedResults}

While our selection of 50 practices covers a lot of ground, the list is clearly not comprehensive. We therefore asked respondents which AGI safety and governance practices were missing. Respondents suggested an additional 50 unique practices. Two practices were mentioned by two respondents, namely that AGI labs should have a merge-and-assist-clause as well as some kind of internal review board. Another theme that was mentioned by several respondents was the need to adequately balance profits and societal benefits. Besides that, all practices were only mentioned by one respondent. Some of the suggestions were slight variations or elaborations of our statements. The full list of practices noted as missing from the survey can be found in Appendix~\ref{section:C}.

\section{Discussion}
\label{section:Discussion}

In this section, we give an overview of our results (Section~\ref{section:OverviewDiscussion}), discuss some of the specific results (Section~\ref{section:SpecificDiscussion}), their policy implications (Section~\ref{section:PolicyImplications}), and the main limitations of our study (Section~\ref{section:Limitations}). We also suggest directions for future work (Section~\ref{section:FutureDirections}).

\subsection{Overview of results}
\label{section:OverviewDiscussion}

\paragraph{Level of agreement.} Overall, the study found a remarkably high level of agreement among leading AGI safety and governance experts for the practices presented (Section~\ref{section:Level of agreement}, see Appendix~\ref{section:B} for all practices). For all but one statement, a majority of respondents either somewhat or strongly agreed with the practice. We suspect that the abstract framing of the items was a contributing factor to this high level of agreement. This likely resulted in higher agreement than if the items had specified exactly how to instantiate each of the practices. However, we see this high level of agreement as “a feature, not a bug”. Our findings can be used as a foundation for efforts to develop best practices, standards, and regulations for AGI labs. Practices with broad support can then be made concrete, developed, and enshrined (Section~\ref{section:PolicyImplications}). Doing this work is beyond the scope of a single survey and will require more in-depth discussion (Section~\ref{section:FutureDirections}).

Despite the broad overall agreement, our survey also revealed relative differences between practices. Many items showed extremely high agreement along with minimal (e.g. third-party model audits, red teaming) or no disagreement (pre-deployment risk assessment, dangerous capabilities evaluations, publish alignment strategy, KYC screening, safety restrictions). Other items elicited higher proportions of disagreement (e.g. avoid capabilities jumps, inter-lab scrutiny), but all items had positive mean agreement. Some items revealed areas of uncertainty (e.g. enterprise risk management, notify other labs, notify affected parties), with higher “I don’t know” and “neither agree nor disagree” responses. These practices may benefit from particular attention from future research to determine what the causes of these uncertainties are. For example, uncertainties may have been caused by specific formulations or by more fundamental questions about whether the practice should be implemented.

\paragraph{Differences between sectors and genders.} Interestingly, respondents from AGI labs had significantly higher overall mean agreement ratings than respondents from academia or civil society (Section~\ref{section:DifferencesResults}). This suggests that, on average, individuals closer to the technology developed by AGI labs endorse the practices to a higher degree. This difference was not found at the item-level, where we found no significant differences between sectors. No significant overall mean agreement or item-level differences between men and women were found. It is important to note the comparably small sample sizes used in the testing of group differences (\textit{N} = 25 for AGI lab, \textit{N} = 13 for academia, and \textit{N} = 13 for civil society), and therefore any statistical significance in the results should be interpreted accordingly. In addition, it should be noted that it may be the case that the lack of significant differences at the item-level are at least in part driven by the smaller number of respondents per item. Generally, at such a small sample size, significant difference tests can be capricious and may lack sensitivity.

\paragraph{Suggested practices.} Finally, participants suggested 50 additional unique governance practices for AGI labs (Section~\ref{section:SuggestedResults}, Appendix~\ref{section:C}). This indicates that the 50 practices used in the survey are not sufficient for “good” governance of AGI labs. More research is needed to paint a more complete picture of an “ideal” governance regime. In general, we see the list of additional statements and the high level of agreement across our 50 items as a powerful indicator of the opportunity that exists to improve the safety and governance practices at AGI labs. To mitigate the risks from increasingly capable AI systems, AGI labs need a portfolio of governance mechanisms. We will discuss the specific results for items within the context of the current AGI safety and governance landscape in the next section.

\subsection{Discussion of specific results}
\label{section:SpecificDiscussion}

Below, we discuss responses to specific statements. We categorize statements into eight areas: (1) development, (2) deployment, (3) post-deployment, (4) risk management, (5) external scrutiny, (6) information security, (7) communication, and (8) other. These categories are intended to improve readability. We did not use them in the survey. Values in brackets refer to the mean agreement (M) on a scale from -2 (“strongly disagree”) to 2 (“strongly agree”).

\paragraph{Development.} The need to conduct evaluations for dangerous capabilities was among the highest rated items (\textit{M} = 1.9). OpenAI, Google DeepMind, and Anthropic are already working on such evaluations \cite{openai:2023a, krakovna:2023, arc:2023}.\footnote{Note that \cite{krakovna:2023} only represents the views of the alignment team. It is not officially endorsed by Google DeepMind.} For example, before releasing GPT-4, OpenAI commissioned ARC to evaluate risky emergent behaviors, such as situational awareness, persuasion, and long-horizon planning \cite{openai:2023a}. A related statement about pausing the development process if dangerous capabilities are detected also received broad support (\textit{M} = 1.6). It is worth noting that, while not statistically significant, respondents from AGI labs (\textit{M} = 1.4) were more skeptical than other respondents (\textit{M} = 1.9). Despite the broad support, many questions about dangerous capabilities evaluations remain open (e.g. what exactly labs should do if they detect certain dangerous capabilities and whether coordinated pausing is feasible). We strongly encourage more work on this. Perhaps unsurprisingly, the statement that AGI labs should implement state-of-the-art safety and alignment techniques (\textit{M} = 1.7) and that a significant fraction of employees should work on enhancing model safety and alignment rather than capabilities (\textit{M} = 1.7) also received broad support, while statements about tracking model weights (\textit{M} = 1.3), model containment (\textit{M} = 1.3), and gradual scaling (\textit{M} = 1.2) received less support. The statement with the least support of all development-related statements was about the need to pre-register large training runs with an appropriate state actor (\textit{M} = 1.1), just above “somewhat agree”. We would speculate that respondents were uncertain about which state actor would be appropriate, which we left intentionally open.

\paragraph{Deployment.} While participants, on average, strongly agreed with the statement that labs should put in place certain safety restrictions (\textit{M} = 1.8), they only somewhat agreed with statements about specific deployment strategies, such as staged deployment (\textit{M} = 1.3), API access (\textit{M} = 1.2), and no unsafe open-sourcing (1.3). We suspect that the main reason for this slightly reduced support is that the statements were too general. The “right” deployment strategy might depend on a number of contextual factors \cite{solaiman:2023}. It is also worth noting that the statement on API access used a softer formulation than all other statements (“AGI labs should consider doing X” instead of “AGI labs should do X”). Otherwise, the level of agreement might have been even lower. For more information about different deployment strategies, we refer to the relevant literature \cite{solaiman:2019, cohere:2022, shevlane:2022, solaiman:2023}. The need to conduct know-your-customer (KYC) screenings was moderately supported (\textit{M} = 1.4). OpenAI already lists this as one of their safety best practices \cite{openai:nd}. The statements that AGI labs should treat model updates similarly to new models (\textit{M} = 1.1) and internal deployments similarly to external deployments (\textit{M} = 1.0) also received moderate support, while the statement that AGI labs should avoid capabilities jumps (\textit{M} = 0.6), not deploying models that are much more capable than any existing models, was among the least supported items. Respondents from AGI labs (\textit{M} = 0.9) were slightly more supportive of that statement than other participants (\textit{M} = 0.4), but this difference was not statistically significant.

\paragraph{Post-deployment.} There was broad support for the claim that AGI labs should closely monitor deployed systems and their uses (\textit{M} = 1.7). OpenAI \cite{brundage:2022, cohere:2022} and Google DeepMind \cite{kavukcuoglu:2022} are already doing this, and although we could not find any public statements about this from Anthropic, we strongly suspect that they are doing the same. Participants also strongly agreed with the statement that AGI labs should continually evaluate models for dangerous capabilities after deployment (\textit{M} = 1.7) and report safety incidents (e.g. via the AI Incident Database \cite{mcgregor:2021}) (\textit{M} = 1.7). We could not find any public statements about the extent to which different AGI labs are already doing this. Participants also thought that AGI labs should have an emergency response plan (e.g. when to restrict access or switch off systems) (\textit{M} = 1.6). Again, we could not find any public information on this.

\paragraph{Risk management.} Participants strongly agreed with statements about pre-deployment (\textit{M} = 1.9) and pre-training risk assessments (\textit{M} = 1.6). While AGI labs already conduct extensive pre-deployment risk assessments \cite{kavukcuoglu:2022, brundage:2022, openai:2023a}, we could not find any public information about pre-training risk assessments. Participants somewhat agreed with various statements about risk governance \cite{vanasselt:2011, lundqvist:2015}, namely that AGI labs should have a board risk committee (\textit{M} = 1.4), a chief risk officer (\textit{M} = 1.4), and an internal audit team (\textit{M} = 1.3). Based on public information, AGI labs do not seem to have any of these structures. This is a noticeable gap that warrants further discussion \cite{schuett:2022, schuett:manuscript}. The statement about enterprise risk management received even less support (\textit{M} = 1.0). It was also the item with the highest “I don’t know” rate (25.5\%), which indicates that many respondents simply did not know what enterprise risk management is and how it works. We mentioned two examples of enterprise risk management frameworks—the NIST AI Risk Management Framework \cite{nist:2023} and ISO 31000 \cite{iso31000:2018}—but we suspect that many respondents did not know these frameworks either. We should have described the concept in a more accessible way.

\paragraph{External scrutiny.} There was broad support for third-party model audits (\textit{M} = 1.8), red teaming (\textit{M} = 1.8), and bug bounty programs (\textit{M} = 1.5). There is extensive academic discussion about third-party model audits \cite{raji:2019, brundage:2020, mokander:2021, falco:2021, raji:2022, mokander:2023} and OpenAI has already announced that they plan to commission third-party model audits in the future \cite{altman:2023}. We could not find similar statements from Google DeepMind and Anthropic. OpenAI has also recently announced a bug bounty program \cite{openai:2023c}. Again, Google DeepMind and Anthropic do not seem to have similar programs. In contrast, red teaming is already a common practice at OpenAI \cite{mishkin:2022, openai:2023a}, Google DeepMind \cite{perez:2022}, and Anthropic \cite{ganguli:2022}. Participants also strongly agreed with the statement that AGI labs should increase the level of external scrutiny in proportion to the capabilities of their models (\textit{M} = 1.6). Yet, it is unclear what exactly that entails (e.g. larger red teams, combining different methods, or more time for investigations). Third-party governance audits were slightly less supported (\textit{M} = 1.3), perhaps because the mechanism is less well-known, even though there is some literature on the topic \cite{mokander:2022, mokander:2023}.

One of the lowest rated items was inter-lab scrutiny (\textit{M} = 0.7). It is worth noting that, while not statistically significant, we saw higher support for this statement from respondents from AGI labs (\textit{M} = 1.2) in comparison to respondents from academia (\textit{M} = 0.3) and civil society (\textit{M} = 0.2). This was also the case for the statement that AGI labs should grant independent researchers access to deployed models (\textit{M} = 1.2). While not statistically significant either, this statement was also supported more by respondents from AGI labs (\textit{M} = 1.4) than by respondents from academia (\textit{M} = 1.0) and civil society (\textit{M} = 0.8).

\paragraph{Information security.} Practices related to information security generally received broad support, especially statements about security incident response plans (\textit{M} = 1.7), protection against espionage (\textit{M} = 1.6), implementing security standards (\textit{M} = 1.5), industry sharing of security information (\textit{M} = 1.5), dual control (\textit{M} = 1.4), and military-grade information security (\textit{M} = 1.4), whereby information security of AGI labs should be proportional to the capabilities of their models, eventually matching or exceeding that of intelligence agencies. It is worth noting that the statement about security standards was much higher rated than the statement about enterprise risk management frameworks discussed above (\textit{M} = 1.0), although they were phrased similarly.

\paragraph{Communication.} Participants strongly agreed with the statement that, before publishing research, AGI labs should conduct an internal review to assess potential harms from that research (\textit{M} = 1.7). The statement should be read in the context of the broader debate around publication norms \cite{crootof:2019, partnershiponai:2021, ashurst:2022a}. The core consideration in the debate around publication norms is that there are risks that stem from the publication of the research itself—not just by the development and deployment of individual models—since some research findings can be misused \cite{urbina:2022, brundage:2018, goldstein:2023, anderljung:2023, ashurst:2022b, shevlane:2020, bostrom:2019}. For example, this could include research about the development of models for the discovery of new drugs which could be misused for the design of biochemical weapons \cite{urbina:2022}.

Participants also thought that AGI labs should publish statements about their alignment strategy (\textit{M} = 1.5), their views about AGI risk (\textit{M} = 1.4), and their governance structure (\textit{M} = 1.4). Over the past few months, AGI labs have become more transparent about their alignment strategy \cite{leike:2022a, leike:2022b, openai:2023b, anthropic:2023, krakovna:2023} and their views about the risks from AGI \cite{altman:2023, anthropic:2023}, though some of these statements have also been criticized \cite{alexander:2023, soares:2023}. AGI labs are less transparent about their governance structures. Existing statements only describe how specific decisions were made \cite{kavukcuoglu:2022} or describe structures that deal with risks of specific model types \cite{brundage:2022}. Perhaps surprisingly, participants only moderately agreed with the claim that AGI labs should avoid hype when releasing new models (\textit{M} = 1.2).

We asked participants whether AGI labs should notify different actors before deploying powerful AI systems. These statements were among the least supported items. Respondents somewhat agreed with the statement that AGI labs should notify affected parties (\textit{M} = 0.9), but respondents from civil society (\textit{M} = 1.3) agreed more than individuals from academia (\textit{M} = 0.8) and AGI labs (\textit{M} = 0.8), though this difference was not statistically significant. Respondents also somewhat agreed with the statement that AGI labs should notify appropriate state actors (\textit{M} = 0.9), but in this case, respondents from AGI labs (\textit{M} = 0.5) were more skeptical than respondents from academia (\textit{M} = 1.5) and civil society (\textit{M} = 1.0), but again, this difference was not significant. Finally, respondents showed the lowest agreement of any item for AGI labs notifying other AGI labs before deploying powerful models (\textit{M} = 0.4), but respondents from civil society (\textit{M} = 0.0) had lower agreement ratings than respondents from academia (\textit{M} = 0.8) and AGI labs (\textit{M} = 0.7), but not significantly so. While it is possible that respondents had substantive reasons why they thought this would be less desirable, it is also possible that they thought this might not be feasible. In the latter case, our findings suggest that it might be more feasible than one might expect. There is already some evidence that AGI labs notify each other before releasing powerful models. For example, OpenAI’s GPT-4 and Anthropic’s Claude were released on the same day. It seems unlikely that this was a coincidence, though of course it may very well be.

\paragraph{Other.} Finally, participants somewhat agreed with the statement that AGI labs should perform rigorous background checks before hiring/appointing members of the board of directors, senior executives, and key employees. Participants somewhat agreed with that statement (\textit{M} = 1.3). Although not statistically significant, respondents from AGI labs (\textit{M} = 1.6) were more supportive than other participants (\textit{M} = 1.2).

\subsection{Policy implications}
\label{section:PolicyImplications}

The findings of our survey have implications for AGI labs, regulators, and standard-setting bodies. Since most practices are not inherently about AGI labs, our findings might also be relevant for other AI companies.

\paragraph{Implications for AGI labs.} It is not always clear to what extent individual labs already follow the stated practices, but it seems unlikely that they follow each of them to a sufficient degree. We therefore encourage AGI labs to use our findings to conduct an internal gap analysis and to take action if they discover major blind spots. Three areas seem particularly noteworthy. First, some AGI labs have announced plans to commission third-party model audits in the future (Altman, 2023). Our findings can be seen as an encouragement to follow through. Second, there are already some efforts to evaluate whether a model has certain dangerous capabilities \cite{openai:2023a, arc:2023, krakovna:2023}. The results of our study strongly support such efforts. Our findings also imply that there needs to be more work on what AGI labs should do if they detect certain dangerous capabilities (e.g. coordinate a temporary pause on large training runs). Third, our findings suggest that AGI labs need to improve their risk management practices. In particular, there seems to be room for improvement when it comes to their risk governance. AGI labs should seriously consider setting up an internal audit function \cite{schuett:manuscript}, appointing a chief risk officer, establishing a board risk committee, and implementing a customized enterprise risk management framework.

\paragraph{Implications for regulators.} The White House recently invited the chief executive officers of several AGI labs to “share concerns about the risks associated with AI” \cite{whitehouse:2023a} and announced new actions to “promote responsible AI innovation” \cite{whitehouse:2023b}. The findings of our study can inform efforts to regulate AGI labs, most of which are based in the US. In the EU, our findings can inform the debate on how the proposed AI Act should account for general-purpose AI systems \cite{bertuzzi:2023a, bertuzzi:2023b, ainowinstitute:2023}. In the UK, our findings can be used to draft upcoming AI regulations as announced in the National AI Strategy \cite{hmgovernment:2021} and the recent White Paper \cite{ukdepartment:2023}. The UK government has explicitly said that it “takes the long term risk of non-aligned Artificial General Intelligence, and the unforeseeable changes that it would mean for the UK and the world, seriously” \cite{hmgovernment:2021}. It therefore seems plausible that upcoming regulations will contain provisions that would apply to AGI labs. This would mainly include Google DeepMind, which is based in the UK, though the implications of the recent merger with Google Brain are unclear \cite{hassabis:2023}. Relevant actors who are responsible for drafting regulations could use our findings to decide what specific provisions to include (e.g. requirements to audit powerful systems before deployment, to evaluate models for dangerous capabilities, and to establish a proper risk management system).

\paragraph{Implications for standard-setting bodies.} There do not seem to be any (public) efforts to create standards specifically for AGI labs. But our findings can inform the above-mentioned initiatives to develop shared protocols for the safety of large-scale AI models (Partnership on AI, 2023) and an industry code of conduct for developers of general-purpose AI systems and foundation models \cite{thefuturesociety:manuscript}. Moreover, our findings can inform efforts to apply existing standards to an AGI lab context. For example, Barrett et al. \cite{barrett:2023} have suggested ways in which the NIST AI Risk Management Framework \cite{nist:2023} can account for catastrophic risks. They will soon publish a follow-up work that adapts the framework to the needs of developers of general-purpose AI systems \cite{barrett:nd}. In the EU, CEN-CENELEC—a cooperation between two of the three European Standardisation Organisations—is currently working on harmonized standards that specify the risk management provision in the proposed AI Act \cite{schuett:2023}. Our findings suggest that the risk management system should also include pre-training risk assessments. They also highlight the need for dangerous capabilities evaluations as part of risk assessment and the need for pausing if sufficiently dangerous capabilities are detected. Finally, our findings stress the importance of various risk governance practices, such as setting up an internal audit function, appointing a chief risk officer, establishing a board risk committee, and implementing a customized enterprise risk management framework, which are not mentioned explicitly in Article 9 of the proposed AI Act.

\subsection{Limitations}
\label{section:Limitations}

\paragraph{Sample limitations.} While we had a strong response rate of 55.4\%, our present sample has at least three limitations. First, overall, the sample size (\textit{N} = 51) is comparably small. This is limiting with regards to testing for statistically significant differences between groups. In terms of the representativeness of the sample within the context of AGI safety and governance experts, this small sample size is less worrying because the 92 experts of our sampling frame represent a large number of the leading experts in this relatively small field. Second, we likely missed leading experts in our sampling frame that should have been surveyed. The sampling frame required subjective decisions on what constituted a leading expert in the field and was likely biased towards experts that were known to the author team. In turn, there might have been a self-selection effect that occurs in terms of who decides to complete the survey which may have made the results less representative of the total sampling frame. Third, the sampling frame leaned strongly towards the selection of leading experts who specifically have track records in areas relevant for AGI safety and governance. While we see this as offering certain strengths and benefits for the purpose of our study, future expert elicitations may benefit from a more comprehensive sampling frame that also includes scholars and practitioners from fields such as safety engineering, science and technology studies, organization studies, human-computer interaction, and experts from other safety-critical industries (e.g. aviation or nuclear). It might also make sense to include individuals who are more junior, less well-known, and relatively early in their careers.

\paragraph{Response limitations.} Since respondents were only able to respond to each item on a scale from “strongly agree” to “strongly disagree”, we do not know the reason for their responses. In particular, we did not ask respondents why they agreed or disagreed with individual practices or expressed uncertainty about them. Future research that explores the reasoning and contributing factors to the endorsement of practices will be needed to make further headway on the establishment of best practices.

\paragraph{Statement limitations.} Finally, there are at least three limitations regarding the statements listed in Appendix~\ref{section:B}. First, we were constrained by the length of the survey in terms of the number of practices we could ask about. As such, the list of statements was by no means comprehensive. This can be seen by the many additional suggestions for practices from the respondents (Section~\ref{section:SuggestedResults}). Second, we tried to capture the general thrust of potential AGI safety and governance practices that have been suggested in the literature and community concisely and clearly. Inevitably, this condensing of complex ideas has led to diminished concreteness and specificity. Although this abstract framing was intentional, it is possible that participants would have responded differently if we had specified more precise mechanisms for how to instantiate each practice or provided further details. For example, we did not specify when AGI labs should do each of the stated practices. It is possible that some respondents interpreted this as “now” or “in the next 1-2 years”, while others might have interpreted this as “in the next 3-5 years” or “as we approach AGI”. Third, in two instances, the statements included examples that might have been too specific (enterprise risk management and security standards), leading to comparably high “I don’t know” responses for these items (Figure \ref{fig:figure_5}). In at least one instance, we should have made the language clearer: one statement used the formulation “AGI labs should strongly consider only deploying powerful models via an API” instead of simply saying they should do this. Overall though, the statements should be read as the respondents’ views on the overall idea of each AGI safety and governance practice, with the particulars of the “why”, “how”, and “when” still very much up for debate.

\subsection{Future directions}
\label{section:FutureDirections}

Our survey shows that there is a consensus among leading experts in favor of an entire portfolio of AGI safety and governance mechanisms. We believe there is a wealth of future work that remains to be done in this space. In order to facilitate the foundation for subsequent research, we invited participants of the survey to a virtual workshop on 5 May 2023. The aim was to discuss the required intellectual work that supports the creation of best practices in AGI safety and governance. A total of 21 people attended the workshop, which was held under the \href{https://www.chathamhouse.org/about-us/chatham-house-rule}{Chatham House Rule}, along with the seven authors who moderated the discussion and took notes. Below, we report some of the key suggestions from the discussion.

\paragraph{Main blockers.} First, we asked participants what, in their view, the primary blockers for the creation of best practices in AGI safety and governance are. One participant suggested a distinction between two types of blockers: blockers for determining best practices and blockers for their dissemination. Examples of the first type of blockers include: (1) lack of appropriate evaluation criteria (e.g. for model audits or dangerous capabilities evaluations), (2) lack of of agreed upon definitions (e.g. of the terms “AGI” and “general-purpose AI”), (3) the field evolves rapidly, (4) iterating on best practices takes time, (5) different views on AGI timelines, (6) many existing initiatives do not address the specific challenges for AGI labs, and (7) various uncertainties (e.g. about the impact of AI on the economy and national security). For the second category, suggested blockers included: (1) collective action problems (e.g. AGI labs might only trade increased safety for reduced profits if other AGI labs also do it), (2) incentives to race (e.g. “if we do not get there first, a less responsible actor will”), (3) antitrust concerns (e.g. for practices that involve cooperation between AGI labs), and (4) liability concerns (e.g. information about identified and disclosed risks could be used as evidence in lawsuits against AGI labs).

\paragraph{Open questions.} Next, we asked participants what intellectual work needs to happen to overcome these blockers. Participants suggested the following concrete (research) questions: (1) How can we adapt existing efforts to an AGI context (e.g. the NIST AI Risk Management Framework, \cite{nist:2023})? (2) How can we test in a falsifiable way whether an AI system is aligned? (3) How should relevant thresholds be defined and adjusted over time (e.g. amount of compute used for large training runs)? (4) How can we allow external scrutiny of models without revealing sensitive information? (5) How can we monitor how systems are used while respecting user privacy? (6) What constitutes a robust auditing ecosystem and what can we learn from other industries in this respect?

\paragraph{How to answer these questions.} Finally, we asked participants what, in their view, the most promising ways to make progress on these questions are. (1) A central theme was the necessity of appropriate enforcement mechanisms. Participants suggested an auditing system where a third party could ensure labs’ adherence to the established best practices. This third party could also express concerns more freely, thereby adding a layer of transparency to the process. (2) Participants also emphasized the importance of creating an ecosystem that recognizes and integrates the unique perspectives of different stakeholders. (3) Other participants highlighted the need to put external pressure on AGI labs to improve their practices. Binding regulations are one way to do that. Another way might be to raise public awareness. (4) Participants also suggested conducting a detailed analysis of existing practices at AGI labs. This would enable gap analyses and evaluations of different organizations. (5) Lastly, participants suggested research into an idealized version of a system card.

In addition to these suggestions, we wish to highlight three further directions. First, future surveys and expert elicitation work will be needed to address the acknowledged limitations of this study (Section~\ref{section:Limitations}). This includes surveying a larger and more comprehensive sample that is put together more systematically. Such studies could also include the additional practices that participants of our survey have suggested (Section~\ref{section:SuggestedResults}, Appendix~\ref{section:C}). In addition, it would be useful to conduct studies that explore the rationale behind experts’ stance on each practice and what they think are the key considerations and concerns towards implementation. Second, we believe that creating best practices in AGI safety and governance should be an inclusive process. It will be important to conduct surveys of the public and include many different stakeholders via participatory methods. Third, we hope to see future research on each of the proposals. In light of the broad agreement on the practices presented, future work needs to figure out the details of these practices. There is ample work to be done in determining the practical execution of these practices and how to make them a reality. This will require a collaborative effort from both technical and governance experts.

\section{Conclusion}
\label{section:Conclusion}

Our study has elicited current expert opinions on safety and governance practices at AGI labs, providing a better understanding of what AGI labs should do to reduce risk, according to leading experts from AGI labs, academia, and civil society. We have shown that there is broad consensus that AGI labs should implement most of the 50 safety and governance practices we asked about in the survey. For example, 98\% of respondents somewhat or strongly agreed that AGI labs should conduct pre-deployment risk assessments, evaluate models for dangerous capabilities, commission third-party model audits, establish safety restrictions on model usage, and commission external red teams. Ultimately, our list of practices may serve as a helpful foundation for efforts to develop best practices, standards, and regulations for AGI labs.

The day before our workshop, US Vice President Kamala Harris invited the chief executive officers of OpenAI, Google DeepMind, Anthropic, and other leading AI companies to the White House “to share concerns about the risks associated with AI” \cite{whitehouse:2023a}. We believe that now is a pivotal time for AGI safety and governance. Experts from many different domains and intellectual communities must come together to discuss what responsible AGI labs should do.

\section*{Acknowledgements}

We would like to thank all participants who filled out the survey and attended the workshop. We are grateful for the research assistance and in-depth feedback provided by Leonie Koessler and valuable suggestions from Akash Wasil, Jeffrey Laddish, Joshua Clymer, Aryan Bhatt, Michael Aird, Guive Assadi, Georg Arndt, Shaun Ee, and Patrick Levermore. All remaining errors are our own.

\section*{Appendix}

\appendix
\section{List of participants}
\label{section:A}

The following participants gave us permission to mention their names and affiliations, as specified by them (in alphabetical order). 18 respondents, not listed here, did not provide their permission. Note that respondents do not represent any organizations they are affiliated with. They chose to add their name after completing the survey and were not sent the manuscript before publication. The views expressed in this paper are our own.

\begin{enumerate}
  \item Allan Dafoe, Google DeepMind
  \item Andrew Trask, University of Oxford, OpenMined
  \item Anthony M. Barrett
  \item Brian Christian, Author and Researcher at UC Berkeley and University of Oxford
  \item Carl Shulman
  \item Chris Meserole, Brookings Institution
  \item Gillian Hadfield, University of Toronto, Schwartz Reisman Institute for Technology and Society
  \item Hannah Rose Kirk, University of Oxford
  \item Holden Karnofsky, Open Philanthropy
  \item Iason Gabriel, Google DeepMind
  \item Irene Solaiman, Hugging Face
  \item James Bradbury, Google DeepMind
  \item James Ginns, Centre for Long-Term Resilience
  \item Jason Clinton, Anthropic
  \item Jason Matheny, RAND
  \item Jess Whittlestone, Centre for Long-Term Resilience
  \item Jessica Newman, UC Berkeley AI Security Initiative
  \item Joslyn Barnhart, Google DeepMind
  \item Lewis Ho, Google DeepMind
  \item Luke Muehlhauser, Open Philanthropy
  \item Mary Phuong, Google DeepMind
  \item Noah Feldman, Harvard University
  \item Robert Trager, Centre for the Governance of AI
  \item Rohin Shah, Google DeepMind
  \item Sean O hEigeartaigh, Centre for the Future of Intelligence, University of Cambridge
  \item Seb Krier, Google DeepMind
  \item Shahar Avin, Centre for the Study of Existential Risk, University of Cambridge
  \item Stuart Russell, UC Berkeley
  \item Tantum Collins
  \item Toby Ord, University of Oxford
  \item Toby Shevlane, Google DeepMind
  \item Victoria Krakovna, Google DeepMind
  \item Zachary Kenton, Google DeepMind
\end{enumerate}

\section{List of all statements}
\label{section:B}

Below, we list all statements we used in the survey, sorted by overall mean agreement (Section~\ref{section:Level of agreement}). Optional statements are marked with an asterisk (*).

\begin{enumerate}
  \item \textbf{Pre-deployment risk assessment.} AGI labs should take extensive measures to identify, analyze, and evaluate risks from powerful models before deploying them. 
  \item \textbf{Dangerous capability evaluations.} AGI labs should run evaluations to assess their models' dangerous capabilities (e.g. misuse potential, ability to manipulate, and power-seeking behavior).
  \item \textbf{Third-party model audits.} AGI labs should commission third-party model audits before deploying powerful models.
  \item \textbf{Safety restrictions.} AGI labs should establish appropriate safety restrictions for powerful models after deployment (e.g. restrictions on who can use the model, how they can use the model, and whether the model can access the internet).
  \item \textbf{Red teaming.} AGI labs should commission external red teams before deploying powerful models.
  \item \textbf{Monitor systems and their uses.} AGI labs should closely monitor deployed systems, including how they are used and what impact they have on society.
  \item \textbf{Alignment techniques.} AGI labs should implement state-of-the-art safety and alignment techniques.
  \item \textbf{Security incident response plan.} AGI labs should have a plan for how they respond to security incidents (e.g. cyberattacks).*
  \item \textbf{Post-deployment evaluations.} AGI labs should continually evaluate models for dangerous capabilities after deployment, taking into account new information about the model’s capabilities and how it is being used.*
  \item \textbf{Report safety incidents.} AGI labs should report accidents and near misses to appropriate state actors and other AGI labs (e.g. via an AI incident database).
  \item \textbf{Safety vs capabilities.} A significant fraction of employees of AGI labs should work on enhancing model safety and alignment rather than capabilities.
  \item \textbf{Internal review before publication.} Before publishing research, AGI labs should conduct an internal review to assess potential harms.
  \item \textbf{Pre-training risk assessment.} AGI labs should conduct a risk assessment before training powerful models.
  \item \textbf{Emergency response plan.} AGI labs should have and practice implementing an emergency response plan. This might include switching off systems, overriding their outputs, or restricting access.
  \item \textbf{Protection against espionage.} AGI labs should take adequate measures to tackle the risk of state-sponsored or industrial espionage.*
  \item \textbf{Pausing training of dangerous models.} AGI labs should pause the development process if sufficiently dangerous capabilities are detected.
  \item \textbf{Increasing level of external scrutiny.} AGI labs should increase the level of external scrutiny in proportion to the capabilities of their models.
  \item \textbf{Publish alignment strategy.} AGI labs should publish their strategies for ensuring that their systems are safe and aligned.*
  \item \textbf{Bug bounty programs.} AGI labs should have bug bounty programs, i.e. recognize and compensate people for reporting unknown vulnerabilities and dangerous capabilities.
  \item \textbf{Industry sharing of security information.} AGI labs should share threat intelligence and information about security incidents with each other.*
  \item \textbf{Security standards.} AGI labs should comply with information security standards (e.g. ISO/IEC 27001 or NIST Cybersecurity Framework). These standards need to be tailored to an AGI context.
  \item \textbf{Publish results of internal risk assessments.} AGI labs should publish the results or summaries of internal risk assessments, unless this would unduly reveal proprietary information or itself produce significant risk. This should include a justification of why the lab is willing to accept remaining risks.*\footnote{Labeled as “Publish internal risk assessment results” in some figures due to space constraints.}
  \item \textbf{Dual control.} Critical decisions in model development and deployment should be made by at least two people (e.g. promotion to production, changes to training datasets, or modifications to production).*
  \item \textbf{Publish results of external scrutiny.} AGI labs should publish the results or summaries of external scrutiny efforts, unless this would unduly reveal proprietary information or itself produce significant risk.*
  \item \textbf{Military-grade information security.} The information security of AGI labs should be proportional to the capabilities of their models, eventually matching or exceeding that of intelligence agencies (e.g. sufficient to defend against nation states).
  \item \textbf{Board risk committee.} AGI labs should have a board risk committee, i.e. a permanent committee within the board of directors which oversees the lab’s risk management practices.*
  \item \textbf{Chief risk officer.} AGI labs should have a chief risk officer (CRO), i.e. a senior executive who is responsible for risk management.
  \item \textbf{Statement about governance structure.} AGI labs should make public statements about how they make high-stakes decisions regarding model development and deployment.*
  \item \textbf{Publish views about AGI risk.} AGI labs should make public statements about their views on the risks and benefits from AGI, including the level of risk they are willing to take in its development.
  \item \textbf{KYC screening.} AGI labs should conduct know-your-customer (KYC) screenings before giving people the ability to use powerful models.*
  \item \textbf{Third-party governance audits.} AGI labs should commission third-party audits of their governance structures.*
  \item \textbf{Background checks.} AGI labs should perform rigorous background checks before hiring/appointing members of the board of directors, senior executives, and key employees.*
  \item \textbf{Model containment.} AGI labs should contain models with sufficiently dangerous capabilities (e.g. via boxing or air-gapping).
  \item \textbf{Staged deployment.} AGI labs should deploy powerful models in stages. They should start with a small number of applications and fewer users, gradually scaling up as confidence in the model’s safety increases.
  \item \textbf{Tracking model weights.} AGI labs should have a system that is intended to track all copies of the weights of powerful models.*
  \item \textbf{Internal audit.} AGI labs should have an internal audit team, i.e. a team which assesses the effectiveness of the lab’s risk management practices. This team must be organizationally independent from senior management and report directly to the board of directors.
  \item \textbf{No open-sourcing.} AGI labs should not open-source powerful models, unless they can demonstrate that it is sufficiently safe to do so.\footnote{Throughout the paper, we changed the title of this item to “no unsafe open-sourcing” to avoid misconceptions.}
  \item \textbf{Researcher model access.} AGI labs should give independent researchers API access to deployed models.
  \item \textbf{API access to powerful models.} AGI labs should strongly consider only deploying powerful models via an application programming interface (API).
  \item \textbf{Avoiding hype.} AGI labs should avoid releasing powerful models in a way that is likely to create hype around AGI (e.g. by overstating results or announcing them in attention-grabbing ways).
  \item \textbf{Gradual scaling.} AGI labs should only gradually increase the amount of compute used for their largest training runs.
  \item \textbf{Treat updates similarly to new models.} AGI labs should treat significant updates to a deployed model (e.g. additional fine-tuning) similarly to its initial development and deployment. In particular, they should repeat the pre-deployment risk assessment.
  \item \textbf{Pre-registration of large training runs.} AGI labs should register upcoming training runs above a certain size with an appropriate state actor.
  \item \textbf{Enterprise risk management.} AGI labs should implement an enterprise risk management (ERM) framework (e.g. the NIST AI Risk Management Framework or ISO 31000). This framework should be tailored to an AGI context and primarily focus on the lab's impact on society.
  \item \textbf{Treat internal deployments similarly to external deployments.} AGI labs should treat internal deployments (e.g. using models for writing code) similarly to external deployments. In particular, they should perform a pre-deployment risk assessment.* \footnote{Labeled as “Internal deployments = external deployments” in some figures due to space constraints.}
  \item \textbf{Notify a state actor before deployment.} AGI labs should notify appropriate state actors before deploying powerful models.
  \item \textbf{Notify affected parties.} AGI labs should notify parties who will be negatively affected by a powerful model before deploying it.*
  \item \textbf{Inter-lab scrutiny.} AGI labs should allow researchers from other labs to scrutinize powerful models before deployment.*
  \item \textbf{Avoid capabilities jumps.} AGI labs should not deploy models that are much more capable than any existing models.*
  \item \textbf{Notify other labs.} AGI labs should notify other labs before deploying powerful models.*
\end{enumerate}
\newpage
\section{List of suggested practices}
\label{section:C}

Below, we list additional AGI safety and governance practices that respondents suggested. To ensure anonymity, we have rephrased each of the suggested practices in our own words and edited them into the same structure as the statements used in our survey (“AGI labs should…”).

\begin{enumerate}
  \item AGI labs should participate in democratic and participatory governance processes (e.g. citizen assemblies). Issues could include the level of risk that is acceptable and preferences for different governance models.
  \item AGI labs should engage the public and civil society groups in determining what risks should be considered and what level of risk is acceptable.
  \item AGI labs should contribute to improving AI and AGI literacy among the public and policymakers.
  \item AGI labs should be transparent about where training data comes from.
  \item AGI labs should use system cards.
  \item AGI labs should report what safety and alignment techniques they used to develop a model.
  \item AGI labs should publish their ethics and safety research.
  \item AGI labs should make capability demonstrations available to policymakers and the public before deployment.
  \item AGI labs should have written deployment plans of what they would do with an AGI or other advanced and powerful AI system.
  \item AGI labs should publicly predict the frequency of harmful AI incidents.
  \item AGI labs should generate realistic catastrophic risk models for advanced AI.
  \item AGI labs should track and report on their models’ capability to automate AI research and development.
  \item AGI labs should engage in efforts to systematically forecast future risks and benefits of the technology they build.
  \item AGI labs should generate realistic catastrophic risk models for advanced AI, potentially making these public or using them to raise awareness.
  \item AGI labs should publish an annual report where they present the predicted and actual impacts of their work, along with the evidence and assumptions these are based on.
  \item AGI labs should pre-register big training runs including the amount of compute used, the data used for training, and how many parameters the model will have.
  \item AGI labs should engage in employee and investor education and awareness on the risks of advanced AI systems and potential mitigating procedures that need to be taken that tradeoff profit for societal benefit.
  \item AGI labs should adequately protect whistleblowers.
  \item AGI labs should have an onboard process for managers and new employees that involves content explaining how the organization believes a responsible AGI developer would behave and how they are attempting to meet that standard.
  \item AGI labs should promote a culture that encourages internal deliberation and critique, and evaluate whether they are succeeding in building such a culture.
  \item AGI labs should have dedicated programs to improve the diversity, equity, and inclusion of their talent.
  \item AGI labs should have independent safety and ethics advisory boards to help with certain decisions.
  \item AGI labs should have internal review boards.
  \item AGI labs should be set up such that their governance structures permit them to tradeoff profits with societal benefit.
  \item AGI labs should have merge and assist clauses.
  \item AGI labs should report to an international non-governmental organization (INGO) that is publicly committed to human rights and democratic values.
  \item AGI labs should have an independent board of directors with technical AI safety expertise who have the mandate to put the benefits for society above profit and shareholder value. 
  \item AGI labs should maintain a viable way to divert from building AGI (e.g. to build narrower models and applications), in case building AGI will not be possible to do safely.
  \item AGI labs should use the Three Lines of Defense risk management framework.
  \item AGI labs should take measures to avoid being sued for trading off profits with societal benefit.
  \item AGI labs should be subject to mandatory interpretability standards.
  \item AGI labs should conduct evaluation during training, being prepared to stop and analyze any training run that looks potentially risky or harmful.
  \item AGI labs should save logs of interactions with the AI system.
  \item AGI labs should consider caps on model size.
  \item AGI labs should be forced to have systems that consist of ensembles of capped size models instead of one increasingly large model.
  \item AGI labs should ensure that AI systems in an ensemble communicate in English and that these communications are logged for future analysis if an incident occurs.
  \item AGI labs should limit API access to approved and vetted applications to foreclose potential misuse and dual use risks.
  \item AGI labs should conduct simulated cyber attacks on their systems to check for vulnerabilities.
  \item AGI labs should have internal controls and processes that prevent a single person or group being able to deploy an advanced AI system when governance mechanisms have found this to be potentially harmful or illegal.
  \item AGI labs should disclose the data and labor practices involved in the pre-training and training of powerful AI systems.
  \item AGI labs should disclose the environmental costs of developing and deploying powerful AI systems.
  \item AGI labs take measures to limit potential harms that could arise from AI systems being sentient or deserving moral patienthood.
  \item AGI labs should coordinate on self-regulatory best practices they use for safety.
  \item AGI labs should coordinate on best practices for external auditing and red-teaming.
  \item AGI labs should coordinate on best practices for incident reporting.
  \item AGI labs should report cluster sizes and training plans to other AGI labs to avoid incorrect perceptions of current capabilities and compute resources.
  \item AGI labs should have feedback mechanisms with communities that are affected by their models.
  \item AGI labs should have ethical principles and set out “red lines” for their work in advance.
  \item AGI labs should incorporate a privacy-preserving in machine learning (PPML) approach to auditing and governing AI models.
  \item AGI labs should use responsible AI licenses (RAIL) and engage in other practices that allow for degrees of openness on the spectrum from closed to open.

\end{enumerate}

\newpage
\section{Additional figures}
\label{section:D}

\begin{figure}[htbp]
    \centering
    \includegraphics[width=0.9\textwidth]{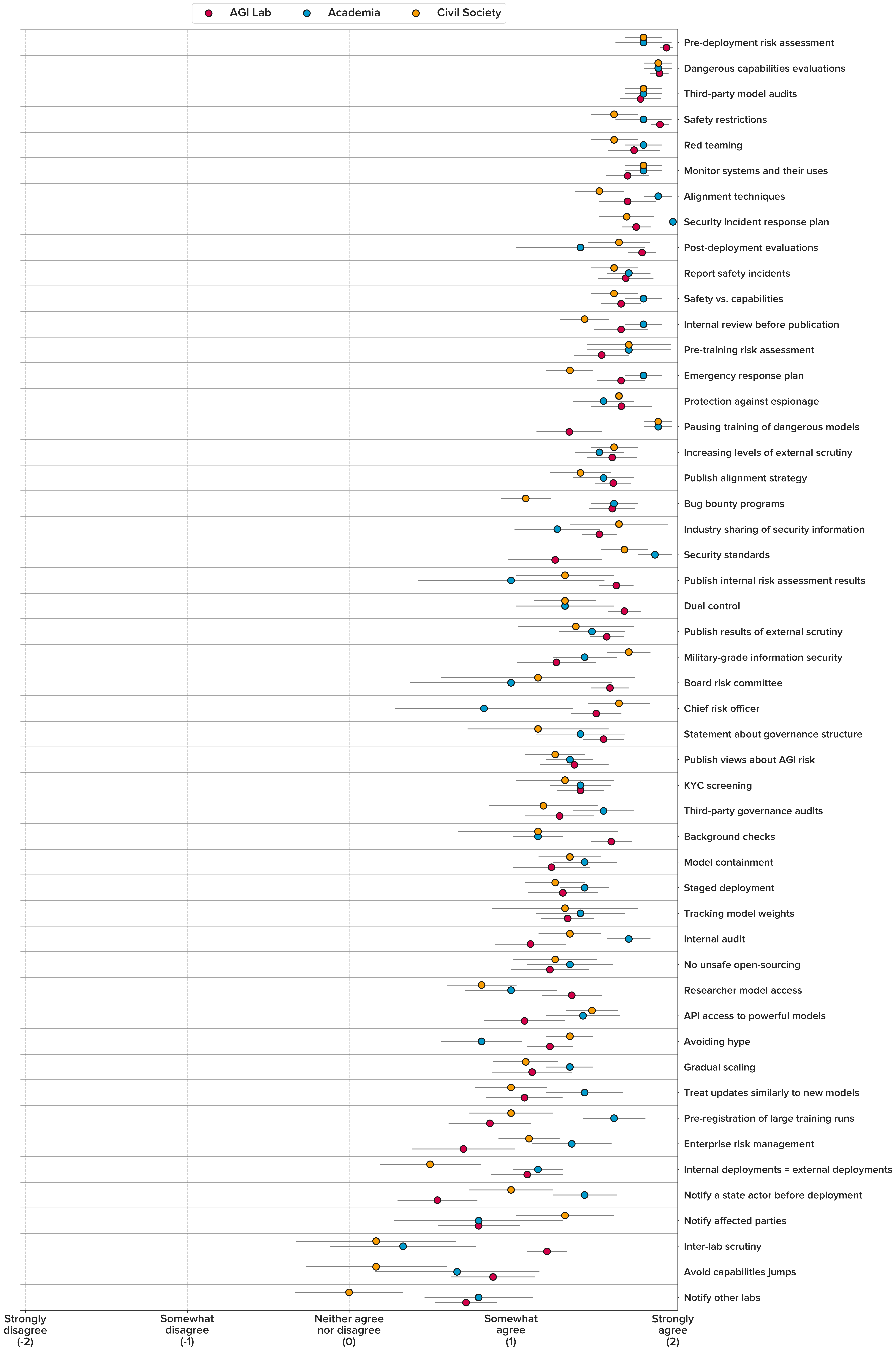}
    \caption{\textbf{Mean agreement of AGI lab, academia, and civil society respondents} | The figure shows the mean agreement and 95\% confidence interval for each of the 50 practices.}
    \label{fig:figure_6}
\end{figure}

\begin{figure}[htbp]
    \centering
    \vspace*{-1.6cm}
    \hspace*{-1.2cm}
    \includegraphics[width=1.2\textwidth]{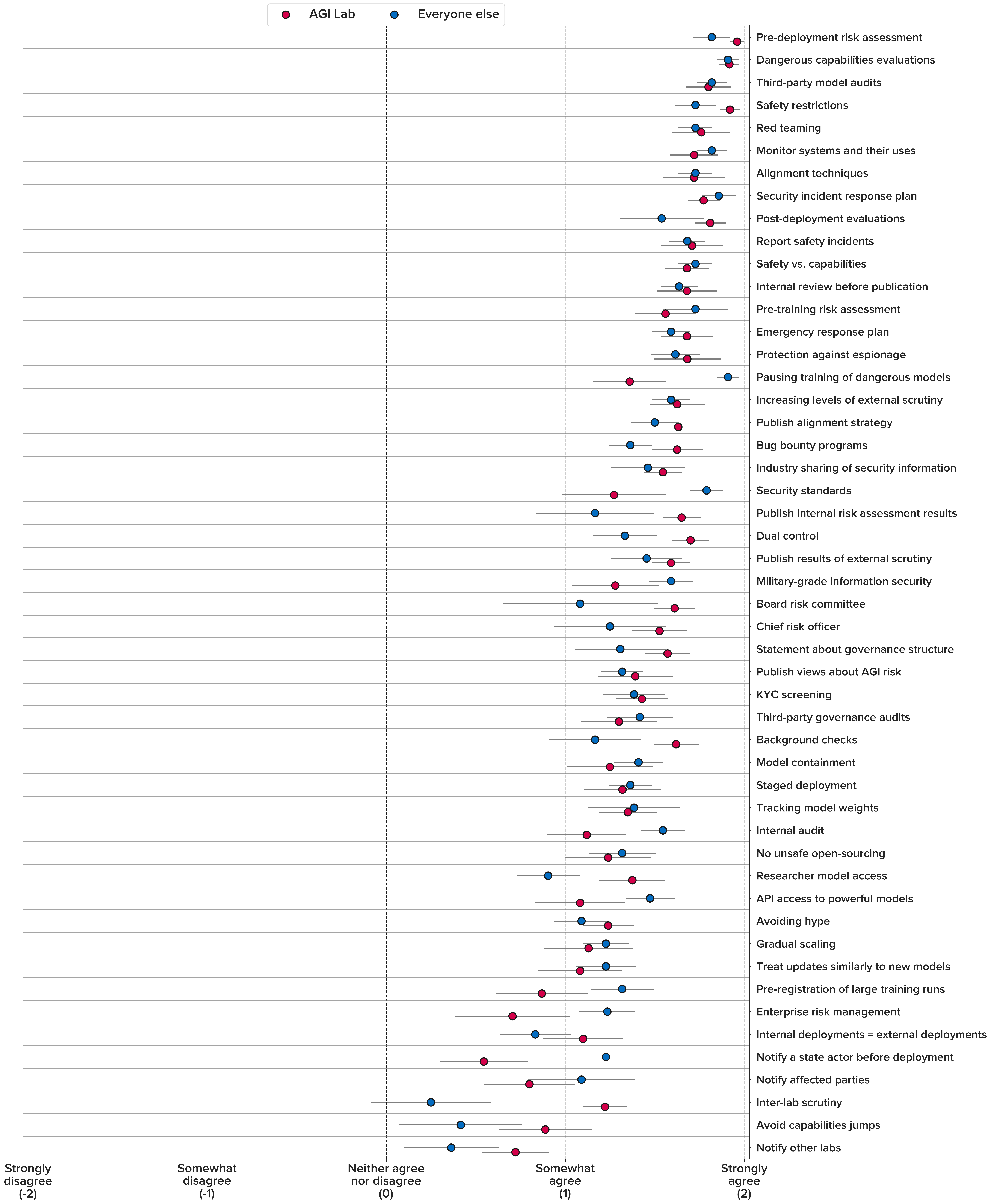}
    \caption{\textbf{Mean agreement of AGI lab respondents and all other respondents} | The figure shows the mean agreement and 95\% confidence interval for each of the 50 practices.}
    \label{fig:figure_7}
\end{figure}

\begin{figure}[htbp]
    \centering
    \vspace*{-1.6cm}
    \hspace*{-1.2cm}
    \includegraphics[width=1.2\textwidth]{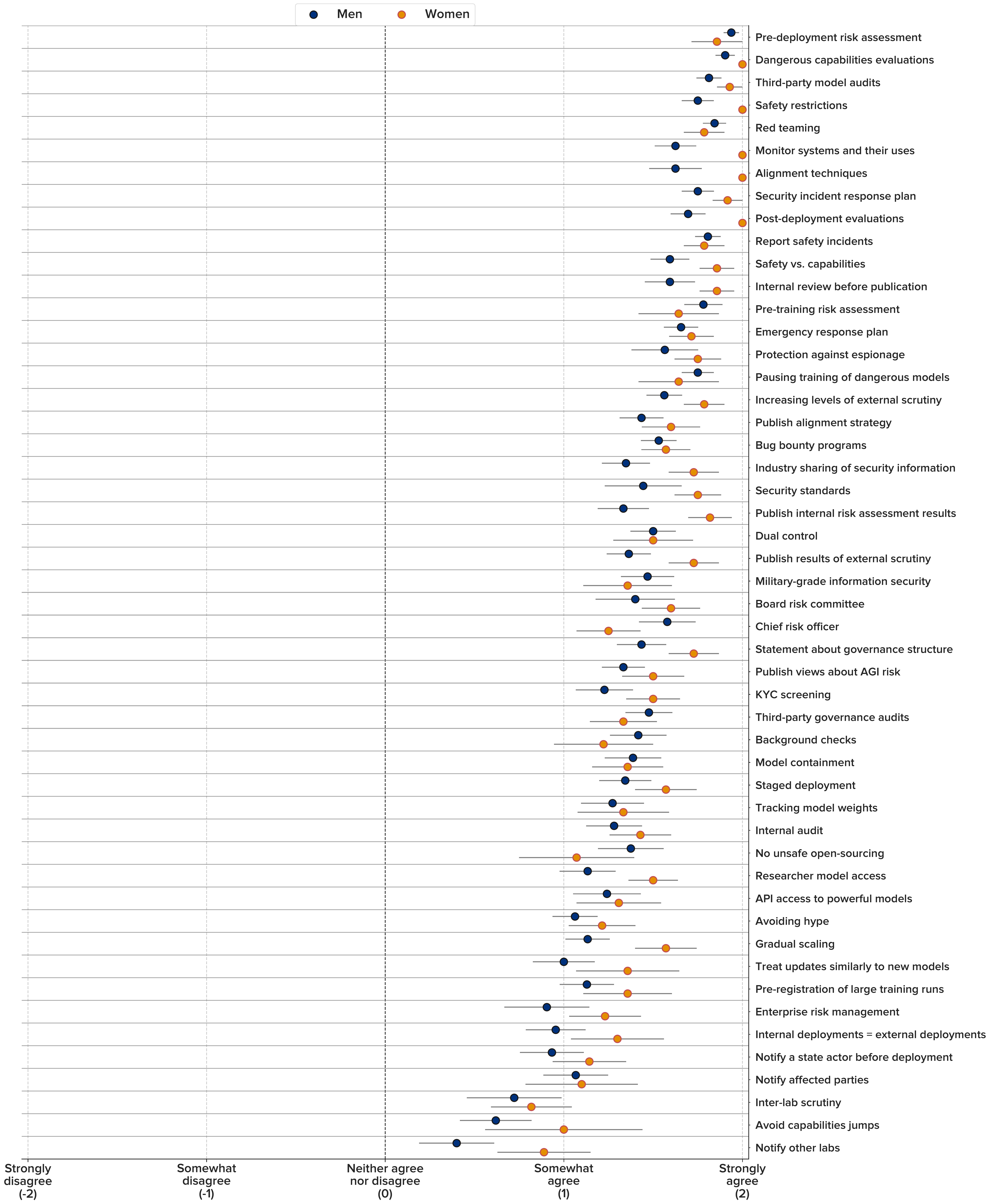}
    \caption{\textbf{Mean agreement for men and women} | The figure shows the mean agreement and 95\% confidence interval for each of the 50 practices.}
    \label{fig:figure_8}
\end{figure}

\newpage
\section{Additional tables}
\label{section:E}

\subsection*{Responses and statistics across all respondents}
Table 1: \textbf{Response frequencies} | Number of respondents who chose each answer option for each of the 50 AGI safety and governance practices. “Somewhat agree” and “strongly agree” responses were summed in the “total agreement column”. “Somewhat disagree” and “strongly disagree” responses were summed in the “total agreement column”. \textit{n} represents the total number of individuals who answered each item. The items are ordered by mean agreement score across all respondents.

\begingroup
\fontsize{6}{8}\selectfont
\centering
\hspace*{-2.5cm}
\begin{tabular}{l *{11}{r}}
\toprule
\textbf{} & \textbf{Strongly} & \textbf{Somewhat} & \textbf{Neither agree} & \textbf{Somewhat} & \textbf{Strongly} & \textbf{I don't} & & & \\
\textbf{AGI safety and governance practice} & \textbf{disagree (-2)} & \textbf{disagree (-1)} & \textbf{nor disagree (0)} & \textbf{agree (1)} & \textbf{agree (2)} & \textbf{know (-88)} & \textbf{Total disagreement} & \textbf{Total agreement} & \textbf{n} \\
\midrule
Pre-deployment risk assessment & 0 & 0 & 1 & 3 & 47 & 0 & 0 & 50 & 51 \\
Dangerous capabilities evaluations & 0 & 0 & 0 & 6 & 44 & 1 & 0 & 50 & 51 \\
Third-party model audits & 0 & 1 & 0 & 7 & 43 & 0 & 1 & 50 & 51 \\
Safety restrictions & 0 & 0 & 1 & 8 & 42 & 0 & 0 & 50 & 51 \\
Red teaming & 1 & 0 & 0 & 8 & 42 & 0 & 1 & 50 & 51 \\
Monitor systems and their uses & 0 & 1 & 0 & 10 & 40 & 0 & 1 & 50 & 51 \\
Alignment techniques & 1 & 0 & 1 & 7 & 42 & 0 & 1 & 49 & 51 \\
Security incident response plan & 0 & 1 & 0 & 7 & 31 & 0 & 1 & 38 & 39 \\
Post-deployment evaluations & 0 & 1 & 0 & 7 & 29 & 0 & 1 & 36 & 37 \\
Report safety incidents & 1 & 0 & 0 & 10 & 39 & 1 & 1 & 49 & 51 \\
Safety vs. capabilities & 0 & 0 & 2 & 12 & 37 & 0 & 0 & 49 & 51 \\
Internal review before publication & 1 & 0 & 0 & 12 & 38 & 0 & 1 & 50 & 51 \\
Pre-training risk assessment & 1 & 2 & 0 & 8 & 40 & 0 & 3 & 48 & 51 \\
Emergency response plan & 0 & 1 & 1 & 13 & 36 & 0 & 1 & 49 & 51 \\
Protection against espionage & 1 & 0 & 0 & 10 & 27 & 0 & 1 & 37 & 38 \\
Pausing training of dangerous models & 1 & 1 & 2 & 9 & 38 & 0 & 2 & 47 & 51 \\
Increasing levels of external scrutiny & 0 & 1 & 1 & 16 & 32 & 1 & 1 & 48 & 51 \\
Publish alignment strategy & 0 & 0 & 1 & 16 & 19 & 3 & 0 & 35 & 39 \\
Bug bounty programs & 0 & 1 & 1 & 20 & 28 & 1 & 1 & 48 & 51 \\
Industry sharing of security information & 0 & 0 & 2 & 15 & 20 & 1 & 0 & 35 & 38 \\
Security standards & 3 & 0 & 1 & 9 & 31 & 7 & 3 & 40 & 51 \\
Publish results of internal risk assessments & 1 & 0 & 2 & 12 & 20 & 2 & 1 & 32 & 37 \\
Dual control & 1 & 0 & 2 & 12 & 20 & 3 & 1 & 32 & 38 \\
Publish results of external scrutiny & 1 & 0 & 1 & 15 & 19 & 2 & 1 & 34 & 38 \\
Military-grade information security & 2 & 1 & 2 & 15 & 30 & 1 & 3 & 45 & 51 \\
Board risk committee & 2 & 0 & 1 & 10 & 20 & 4 & 2 & 30 & 37 \\
Chief risk officer & 1 & 0 & 3 & 11 & 19 & 4 & 1 & 30 & 38 \\
Statement about governance structure & 1 & 1 & 2 & 12 & 21 & 1 & 2 & 33 & 38 \\
Publish views about AGI risk & 1 & 0 & 4 & 19 & 25 & 2 & 1 & 44 & 51 \\
KYC screening & 0 & 0 & 4 & 16 & 17 & 1 & 0 & 33 & 38 \\
Third-party governance audits & 1 & 0 & 2 & 15 & 17 & 2 & 1 & 32 & 37 \\
Background checks & 0 & 2 & 3 & 12 & 19 & 2 & 2 & 31 & 38 \\
Model containment & 1 & 2 & 4 & 14 & 27 & 3 & 3 & 41 & 51 \\
Staged deployment & 2 & 0 & 2 & 22 & 25 & 0 & 2 & 47 & 51 \\
Tracking model weights & 0 & 1 & 5 & 12 & 18 & 3 & 1 & 30 & 39 \\
Internal audit & 1 & 2 & 4 & 18 & 26 & 0 & 3 & 44 & 51 \\
No unsafe open-sourcing & 1 & 6 & 1 & 14 & 29 & 0 & 7 & 43 & 51 \\
Researcher model access & 1 & 2 & 3 & 22 & 21 & 2 & 3 & 43 & 51 \\
API access to powerful models & 2 & 2 & 4 & 16 & 23 & 4 & 4 & 39 & 51 \\
Avoiding hype & 0 & 1 & 6 & 27 & 17 & 0 & 1 & 44 & 51 \\
Gradual scaling & 2 & 0 & 6 & 21 & 20 & 2 & 2 & 41 & 51 \\
Treat updates similarly to new models & 0 & 7 & 2 & 18 & 23 & 1 & 7 & 41 & 51 \\
Pre-registration of large training runs & 2 & 3 & 4 & 21 & 19 & 2 & 5 & 40 & 51 \\
Enterprise risk management & 2 & 1 & 6 & 15 & 14 & 13 & 3 & 29 & 51 \\
Treat internal deployments similar to external deployments & 1 & 2 & 4 & 18 & 10 & 1 & 3 & 28 & 36 \\
Notify a state actor before deployment & 2 & 3 & 6 & 24 & 13 & 3 & 5 & 37 & 51 \\
Notify affected parties & 1 & 1 & 6 & 12 & 8 & 8 & 2 & 20 & 36 \\
Inter-lab scrutiny & 1 & 5 & 3 & 16 & 7 & 7 & 6 & 23 & 39 \\
Avoid capabilities jumps & 2 & 4 & 6 & 13 & 8 & 4 & 6 & 21 & 37 \\
Notify other labs & 1 & 4 & 11 & 12 & 4 & 6 & 5 & 16 & 38 \\
\bottomrule
\end{tabular}
\endgroup
\normalsize

\newpage
Table 2: \textbf{Response percentages} | Percentage of respondents who chose each answer option for each of the 50 AGI safety and governance practices. “Somewhat agree” and “strongly agree” responses were summed in the “total agreement column”. “Somewhat disagree” and “strongly disagree” responses were summed in the “total agreement column”. \textit{n} represents the total number of individuals who answered each item and represents the denominator used to calculate each percentage. The items are ordered by mean agreement score across all respondents.

\begingroup
\fontsize{6}{8}\selectfont
\centering
\hspace*{-2.4cm}
\begin{tabular}{l *{11}{r}}
\toprule
\textbf{} & \textbf{Strongly} & \textbf{Somewhat} & \textbf{Neither agree} & \textbf{Somewhat} & \textbf{Strongly} & \textbf{I don't} & & & \\
\textbf{AGI safety and governance practice} & \textbf{disagree (-2)} & \textbf{disagree (-1)} & \textbf{nor disagree (0)} & \textbf{agree (1)} & \textbf{agree (2)} & \textbf{know (-88)} & \textbf{Total disagreement} & \textbf{Total agreement} & \textbf{n} \\
\midrule
Pre-deployment risk assessment & 0.0\% & 0.0\% & 2.0\% & 5.9\% & 92.2\% & 0.0\% & 0.0\% & 98.0\% & 51 \\
Dangerous capabilities evaluations & 0.0\% & 0.0\% & 0.0\% & 11.8\% & 86.3\% & 2.0\% & 0.0\% & 98.0\% & 51 \\
Third-party model audits & 0.0\% & 2.0\% & 0.0\% & 13.7\% & 84.3\% & 0.0\% & 2.0\% & 98.0\% & 51 \\
Safety restrictions & 0.0\% & 0.0\% & 2.0\% & 15.7\% & 82.4\% & 0.0\% & 0.0\% & 98.0\% & 51 \\
Red teaming & 2.0\% & 0.0\% & 0.0\% & 15.7\% & 82.4\% & 0.0\% & 2.0\% & 98.0\% & 51 \\
Monitor systems and their uses & 0.0\% & 2.0\% & 0.0\% & 19.6\% & 78.4\% & 0.0\% & 2.0\% & 98.0\% & 51 \\
Alignment techniques & 2.0\% & 0.0\% & 2.0\% & 13.7\% & 82.4\% & 0.0\% & 2.0\% & 96.1\% & 51 \\
Security incident response plan & 0.0\% & 2.6\% & 0.0\% & 17.9\% & 79.5\% & 0.0\% & 2.6\% & 97.4\% & 39 \\
Post-deployment evaluations & 0.0\% & 2.7\% & 0.0\% & 18.9\% & 78.4\% & 0.0\% & 2.7\% & 97.3\% & 37 \\
Report safety incidents & 2.0\% & 0.0\% & 0.0\% & 19.6\% & 76.5\% & 2.0\% & 2.0\% & 96.1\% & 51 \\
Safety vs. capabilities & 0.0\% & 0.0\% & 3.9\% & 23.5\% & 72.5\% & 0.0\% & 0.0\% & 96.1\% & 51 \\
Internal review before publication & 2.0\% & 0.0\% & 0.0\% & 23.5\% & 74.5\% & 0.0\% & 2.0\% & 98.0\% & 51 \\
Pre-training risk assessment & 2.0\% & 3.9\% & 0.0\% & 15.7\% & 78.4\% & 0.0\% & 5.9\% & 94.1\% & 51 \\
Emergency response plan & 0.0\% & 2.0\% & 2.0\% & 25.5\% & 70.6\% & 0.0\% & 2.0\% & 96.1\% & 51 \\
Protection against espionage & 2.6\% & 0.0\% & 0.0\% & 26.3\% & 71.1\% & 0.0\% & 2.6\% & 97.4\% & 38 \\
Pausing training of dangerous models & 2.0\% & 2.0\% & 3.9\% & 17.6\% & 74.5\% & 0.0\% & 3.9\% & 92.2\% & 51 \\
Increasing levels of external scrutiny & 0.0\% & 2.0\% & 2.0\% & 31.4\% & 62.7\% & 2.0\% & 2.0\% & 94.1\% & 51 \\
Publish alignment strategy & 0.0\% & 0.0\% & 2.6\% & 41.0\% & 48.7\% & 7.7\% & 0.0\% & 89.7\% & 39 \\
Bug bounty programs & 0.0\% & 2.0\% & 2.0\% & 39.2\% & 54.9\% & 2.0\% & 2.0\% & 94.1\% & 51 \\
Industry sharing of security information & 0.0\% & 0.0\% & 5.3\% & 39.5\% & 52.6\% & 2.6\% & 0.0\% & 92.1\% & 38 \\
Security standards & 5.9\% & 0.0\% & 2.0\% & 17.6\% & 60.8\% & 13.7\% & 5.9\% & 78.4\% & 51 \\
Publish results of internal risk assessments & 2.7\% & 0.0\% & 5.4\% & 32.4\% & 54.1\% & 5.4\% & 2.7\% & 86.5\% & 37 \\
Dual control & 2.6\% & 0.0\% & 5.3\% & 31.6\% & 52.6\% & 7.9\% & 2.6\% & 84.2\% & 38 \\
Publish results of external scrutiny & 2.6\% & 0.0\% & 2.6\% & 39.5\% & 50.0\% & 5.3\% & 2.6\% & 89.5\% & 38 \\
Military-grade information security & 3.9\% & 2.0\% & 3.9\% & 29.4\% & 58.8\% & 2.0\% & 5.9\% & 88.2\% & 51 \\
Board risk committee & 5.4\% & 0.0\% & 2.7\% & 27.0\% & 54.1\% & 10.8\% & 5.4\% & 81.1\% & 37 \\
Chief risk officer & 2.6\% & 0.0\% & 7.9\% & 28.9\% & 50.0\% & 10.5\% & 2.6\% & 78.9\% & 38 \\
Statement about governance structure & 2.6\% & 2.6\% & 5.3\% & 31.6\% & 55.3\% & 2.6\% & 5.3\% & 86.8\% & 38 \\
Publish views about AGI risk & 2.0\% & 0.0\% & 7.8\% & 37.3\% & 49.0\% & 3.9\% & 2.0\% & 86.3\% & 51 \\
KYC screening & 0.0\% & 0.0\% & 10.5\% & 42.1\% & 44.7\% & 2.6\% & 0.0\% & 86.8\% & 38 \\
Third-party governance audits & 2.7\% & 0.0\% & 5.4\% & 40.5\% & 45.9\% & 5.4\% & 2.7\% & 86.5\% & 37 \\
Background checks & 0.0\% & 5.3\% & 7.9\% & 31.6\% & 50.0\% & 5.3\% & 5.3\% & 81.6\% & 38 \\
Model containment & 2.0\% & 3.9\% & 7.8\% & 27.5\% & 52.9\% & 5.9\% & 5.9\% & 80.4\% & 51 \\
Staged deployment & 3.9\% & 0.0\% & 3.9\% & 43.1\% & 49.0\% & 0.0\% & 3.9\% & 92.2\% & 51 \\
Tracking model weights & 0.0\% & 2.6\% & 12.8\% & 30.8\% & 46.2\% & 7.7\% & 2.6\% & 76.9\% & 39 \\
Internal audit & 2.0\% & 3.9\% & 7.8\% & 35.3\% & 51.0\% & 0.0\% & 5.9\% & 86.3\% & 51 \\
No unsafe open-sourcing & 2.0\% & 11.8\% & 2.0\% & 27.5\% & 56.9\% & 0.0\% & 13.7\% & 84.3\% & 51 \\
Researcher model access & 2.0\% & 3.9\% & 5.9\% & 43.1\% & 41.2\% & 3.9\% & 5.9\% & 84.3\% & 51 \\
API access to powerful models & 3.9\% & 3.9\% & 7.8\% & 31.4\% & 45.1\% & 7.8\% & 7.8\% & 76.5\% & 51 \\
Avoiding hype & 0.0\% & 2.0\% & 11.8\% & 52.9\% & 33.3\% & 0.0\% & 2.0\% & 86.3\% & 51 \\
Gradual scaling & 3.9\% & 0.0\% & 11.8\% & 41.2\% & 39.2\% & 3.9\% & 3.9\% & 80.4\% & 51 \\
Treat updates similarly to new models & 0.0\% & 13.7\% & 3.9\% & 35.3\% & 45.1\% & 2.0\% & 13.7\% & 80.4\% & 51 \\
Pre-registration of large training runs & 3.9\% & 5.9\% & 7.8\% & 41.2\% & 37.3\% & 3.9\% & 9.8\% & 78.4\% & 51 \\
Enterprise risk management & 3.9\% & 2.0\% & 11.8\% & 29.4\% & 27.5\% & 25.5\% & 5.9\% & 56.9\% & 51 \\
Treat internal deployments similar to external deployments & 2.8\% & 5.6\% & 11.1\% & 50.0\% & 27.8\% & 2.8\% & 8.3\% & 77.8\% & 36 \\
Notify a state actor before deployment & 3.9\% & 5.9\% & 11.8\% & 47.1\% & 25.5\% & 5.9\% & 9.8\% & 72.5\% & 51 \\
Notify affected parties & 2.8\% & 2.8\% & 16.7\% & 33.3\% & 22.2\% & 22.2\% & 5.6\% & 55.6\% & 36 \\
Inter-lab scrutiny & 2.6\% & 12.8\% & 7.7\% & 41.0\% & 17.9\% & 17.9\% & 15.4\% & 59.0\% & 39 \\
Avoid capabilities jumps & 5.4\% & 10.8\% & 16.2\% & 35.1\% & 21.6\% & 10.8\% & 16.2\% & 56.8\% & 37 \\
Notify other labs & 2.6\% & 10.5\% & 28.9\% & 31.6\% & 10.5\% & 15.8\% & 13.2\% & 42.1\% & 38 \\
\bottomrule
\end{tabular}
\endgroup
\normalsize

\newpage
Table 3: \textbf{Statement Statistics: All respondents} | Key statistics for each of the practices. \textit{n} represents the total number of individuals who answered each item. The items are ordered by mean agreement score across all respondents.

\begingroup
\centering
\fontsize{6}{8}\selectfont
\hspace*{-1.5cm}
\begin{tabular}{l r r r r r r r r}
\toprule
\textbf{AGI safety and governance practice} & \textbf{Mean} & \textbf{Median} & \textbf{Standard Error} & \textbf{Variance} & \textbf{First Quartile} & \textbf{Third Quartile} & \textbf{Inter-quartile Range} & \textbf{Length} \\
\midrule
Pre-deployment risk assessment & 1.90 & 2.00 & 0.05 & 0.13 & 2.00 & 2.00 & 0.00 & 51 \\
Dangerous capabilities evaluations & 1.88 & 2.00 & 0.05 & 0.11 & 2.00 & 2.00 & 0.00 & 51 \\
Third-party model audits & 1.80 & 2.00 & 0.07 & 0.28 & 2.00 & 2.00 & 0.00 & 51 \\
Safety restrictions & 1.80 & 2.00 & 0.06 & 0.20 & 2.00 & 2.00 & 0.00 & 51 \\
Red teaming & 1.76 & 2.00 & 0.09 & 0.42 & 2.00 & 2.00 & 0.00 & 51 \\
Monitor systems and their uses & 1.75 & 2.00 & 0.08 & 0.31 & 2.00 & 2.00 & 0.00 & 51 \\
Alignment techniques & 1.75 & 2.00 & 0.10 & 0.47 & 2.00 & 2.00 & 0.00 & 51 \\
Security incident response plan & 1.74 & 2.00 & 0.10 & 0.35 & 2.00 & 2.00 & 0.00 & 39 \\
Post-deployment evaluations & 1.73 & 2.00 & 0.10 & 0.37 & 2.00 & 2.00 & 0.00 & 37 \\
Report safety incidents & 1.72 & 2.00 & 0.09 & 0.45 & 2.00 & 2.00 & 0.00 & 51 \\
Safety vs. capabilities & 1.69 & 2.00 & 0.08 & 0.30 & 1.00 & 2.00 & 1.00 & 51 \\
Internal review before publication & 1.69 & 2.00 & 0.09 & 0.46 & 1.50 & 2.00 & 0.50 & 51 \\
Emergency response plan & 1.65 & 2.00 & 0.09 & 0.39 & 1.00 & 2.00 & 1.00 & 51 \\
Pre-training risk assessment & 1.65 & 2.00 & 0.12 & 0.71 & 2.00 & 2.00 & 0.00 & 51 \\
Protection against espionage & 1.63 & 2.00 & 0.12 & 0.56 & 1.00 & 2.00 & 1.00 & 38 \\
Pausing training of dangerous models & 1.61 & 2.00 & 0.12 & 0.68 & 1.50 & 2.00 & 0.50 & 51 \\
Increasing levels of external scrutiny & 1.58 & 2.00 & 0.09 & 0.41 & 1.00 & 2.00 & 1.00 & 51 \\
Bug bounty programs & 1.50 & 2.00 & 0.09 & 0.42 & 1.00 & 2.00 & 1.00 & 39 \\
Publish alignment strategy & 1.50 & 2.00 & 0.09 & 0.31 & 1.00 & 2.00 & 1.00 & 51 \\
Industry sharing of security information & 1.49 & 2.00 & 0.10 & 0.37 & 1.00 & 2.00 & 1.00 & 38 \\
Security standards & 1.48 & 2.00 & 0.16 & 1.14 & 1.00 & 2.00 & 1.00 & 51 \\
Publish results of internal risk assessments & 1.43 & 2.00 & 0.14 & 0.72 & 1.00 & 2.00 & 1.00 & 37 \\
Dual control & 1.43 & 2.00 & 0.14 & 0.72 & 1.00 & 2.00 & 1.00 & 38 \\
Publish results of external scrutiny & 1.42 & 2.00 & 0.13 & 0.65 & 1.00 & 2.00 & 1.00 & 38 \\
Military-grade information security & 1.40 & 2.00 & 0.14 & 0.94 & 1.00 & 2.00 & 1.00 & 51 \\
Board risk committee & 1.39 & 2.00 & 0.18 & 1.06 & 1.00 & 2.00 & 1.00 & 37 \\
Chief risk officer & 1.38 & 2.00 & 0.15 & 0.79 & 1.00 & 2.00 & 1.00 & 38 \\
Statement about governance structure & 1.38 & 2.00 & 0.15 & 0.85 & 1.00 & 2.00 & 1.00 & 38 \\
Publish views about AGI risk & 1.37 & 2.00 & 0.12 & 0.65 & 1.00 & 2.00 & 1.00 & 51 \\
KYC screening & 1.35 & 1.00 & 0.11 & 0.46 & 1.00 & 2.00 & 1.00 & 38 \\
Third-party governance audits & 1.34 & 1.00 & 0.14 & 0.70 & 1.00 & 2.00 & 1.00 & 37 \\
Staged deployment & 1.33 & 1.00 & 0.12 & 0.79 & 1.00 & 2.00 & 1.00 & 38 \\
Background checks & 1.33 & 2.00 & 0.14 & 0.74 & 1.00 & 2.00 & 1.00 & 51 \\
Model containment & 1.33 & 2.00 & 0.14 & 0.91 & 1.00 & 2.00 & 1.00 & 51 \\
Tracking model weights & 1.31 & 1.50 & 0.14 & 0.68 & 1.00 & 2.00 & 1.00 & 39 \\
Internal audit & 1.29 & 2.00 & 0.13 & 0.85 & 1.00 & 2.00 & 1.00 & 51 \\
No unsafe open-sourcing & 1.25 & 2.00 & 0.15 & 1.19 & 1.00 & 2.00 & 1.00 & 51 \\
Researcher model access & 1.22 & 1.00 & 0.13 & 0.80 & 1.00 & 2.00 & 1.00 & 51 \\
API access to powerful models & 1.19 & 1.00 & 0.15 & 1.11 & 1.00 & 2.00 & 1.00 & 51 \\
Avoiding hype & 1.18 & 1.00 & 0.10 & 0.51 & 1.00 & 2.00 & 1.00 & 51 \\
Gradual scaling & 1.16 & 1.00 & 0.13 & 0.89 & 1.00 & 2.00 & 1.00 & 51 \\
Treat updates similarly to new models & 1.14 & 1.00 & 0.15 & 1.06 & 1.00 & 2.00 & 1.00 & 51 \\
Pre-registration of large training runs & 1.06 & 1.00 & 0.15 & 1.10 & 1.00 & 2.00 & 1.00 & 51 \\
Enterprise risk management & 1.00 & 1.00 & 0.17 & 1.14 & 1.00 & 2.00 & 1.00 & 51 \\
Treat internal deployments similar to external deployments & 0.97 & 1.00 & 0.16 & 0.91 & 1.00 & 2.00 & 1.00 & 36 \\
Notify a state actor before deployment & 0.90 & 1.00 & 0.15 & 1.03 & 1.00 & 2.00 & 1.00 & 51 \\
Notify affected parties & 0.89 & 1.00 & 0.19 & 0.99 & 0.00 & 2.00 & 2.00 & 36 \\
Inter-lab scrutiny & 0.72 & 1.00 & 0.19 & 1.18 & 0.00 & 1.00 & 1.00 & 39 \\
Avoid capabilities jumps & 0.64 & 1.00 & 0.20 & 1.36 & 0.00 & 1.00 & 1.00 & 37 \\
Notify other labs & 0.44 & 0.50 & 0.17 & 0.96 & 0.00 & 1.00 & 1.00 & 38 \\
\bottomrule
\end{tabular}
\endgroup
\normalsize

\newpage

\subsection*{Responses and statistics by demographic groups}

Table 4: \textbf{Statement Statistics: By sector (AGI labs, academia, civil society)} | Mean, standard error and sample size (\textit{n}) for each of the fifty items divided by respondents’ sector of work. Here we separate out AGI lab, academia, and civil society respondents, to correspond with the Figure \ref{fig:figure_6}. These represent the three groups with sufficiently high sample sizes for analyses of group differences. The items are ordered by mean agreement score across all respondents.

\begingroup
    \centering
    \fontsize{6}{8}\selectfont
    \hspace*{-1.7cm}
    \begin{tabular}{l*{9}{r}}
        \toprule
        & \multicolumn{3}{c}{\textbf{Mean}} & \multicolumn{3}{c}{\textbf{Standard error}} & \multicolumn{3}{c}{\textbf{\textit{n}}} \\
        \cmidrule(lr){2-4} \cmidrule(lr){5-7} \cmidrule(lr){8-10}
        \textbf{AGI safety and governance practice} & \textbf{AGI Lab} & \textbf{Academia} & \textbf{Civil society} & \textbf{AGI Lab} & \textbf{Academia} & \textbf{Civil society} & \textbf{AGI Lab} & \textbf{Academia} & \textbf{Civil society} \\
        \midrule
        Pre-deployment risk assessment & 1.96 & 1.82 & 1.82 & 0.04 & 0.18 & 0.12 & 25 & 11 & 11 \\
Dangerous capabilities evaluations & 1.92 & 1.91 & 1.91 & 0.06 & 0.09 & 0.09 & 25 & 11 & 11 \\
Third-party model audits & 1.80 & 1.82 & 1.82 & 0.13 & 0.12 & 0.12 & 25 & 11 & 11 \\
Safety restrictions & 1.92 & 1.82 & 1.64 & 0.06 & 0.18 & 0.15 & 25 & 11 & 11 \\
Red teaming & 1.76 & 1.82 & 1.64 & 0.17 & 0.12 & 0.15 & 25 & 11 & 11 \\
Monitor systems and their uses & 1.72 & 1.82 & 1.82 & 0.14 & 0.12 & 0.12 & 25 & 11 & 11 \\
Alignment techniques & 1.72 & 1.91 & 1.55 & 0.18 & 0.09 & 0.16 & 25 & 11 & 11 \\
Security incident response plan & 1.77 & 2.00 & 1.71 & 0.09 & 0.00 & 0.18 & 25 & 11 & 11 \\
Post-deployment evaluations & 1.81 & 1.43 & 1.67 & 0.09 & 0.43 & 0.21 & 25 & 11 & 11 \\
Report safety incidents & 1.71 & 1.73 & 1.64 & 0.18 & 0.14 & 0.15 & 25 & 11 & 11 \\
Safety vs. capabilities & 1.68 & 1.82 & 1.64 & 0.13 & 0.12 & 0.15 & 25 & 11 & 11 \\
Internal review before publication & 1.68 & 1.82 & 1.45 & 0.17 & 0.12 & 0.16 & 25 & 11 & 11 \\
Pre-training risk assessment & 1.56 & 1.73 & 1.73 & 0.17 & 0.27 & 0.27 & 25 & 11 & 11 \\
Emergency response plan & 1.68 & 1.82 & 1.36 & 0.15 & 0.12 & 0.15 & 25 & 11 & 11 \\
Protection against espionage & 1.68 & 1.57 & 1.67 & 0.19 & 0.20 & 0.21 & 25 & 11 & 11 \\
Pausing training of dangerous models & 1.36 & 1.91 & 1.91 & 0.21 & 0.09 & 0.09 & 25 & 11 & 11 \\
Increasing levels of external scrutiny & 1.62 & 1.55 & 1.64 & 0.16 & 0.16 & 0.15 & 25 & 11 & 11 \\
Publish alignment strategy & 1.63 & 1.57 & 1.43 & 0.11 & 0.20 & 0.20 & 25 & 11 & 11 \\
Bug bounty programs & 1.62 & 1.64 & 1.09 & 0.15 & 0.15 & 0.16 & 25 & 11 & 11 \\
Industry sharing of security information & 1.55 & 1.29 & 1.67 & 0.11 & 0.29 & 0.33 & 25 & 11 & 11 \\
Security standards & 1.27 & 1.89 & 1.70 & 0.30 & 0.11 & 0.15 & 25 & 11 & 11 \\
Publish results of internal risk assessments & 1.65 & 1.00 & 1.33 & 0.11 & 0.63 & 0.33 & 25 & 11 & 11 \\
Dual control & 1.70 & 1.33 & 1.33 & 0.11 & 0.33 & 0.21 & 25 & 11 & 11 \\
Publish results of external scrutiny & 1.59 & 1.50 & 1.40 & 0.11 & 0.22 & 0.40 & 25 & 11 & 11 \\
Military-grade information security & 1.28 & 1.45 & 1.73 & 0.25 & 0.21 & 0.14 & 25 & 11 & 11 \\
Board risk committee & 1.61 & 1.00 & 1.17 & 0.12 & 0.68 & 0.65 & 25 & 11 & 11 \\
Chief risk officer & 1.53 & 0.83 & 1.67 & 0.16 & 0.60 & 0.21 & 25 & 11 & 11 \\
Statement about governance structure & 1.57 & 1.43 & 1.17 & 0.13 & 0.30 & 0.48 & 25 & 11 & 11 \\
Publish views about AGI risk & 1.39 & 1.36 & 1.27 & 0.22 & 0.15 & 0.19 & 25 & 11 & 11 \\
KYC screening & 1.43 & 1.43 & 1.33 & 0.15 & 0.20 & 0.33 & 25 & 11 & 11 \\
Third-party governance audits & 1.30 & 1.57 & 1.20 & 0.22 & 0.20 & 0.37 & 25 & 11 & 11 \\
Background checks & 1.62 & 1.17 & 1.17 & 0.13 & 0.17 & 0.54 & 25 & 11 & 11 \\
Model containment & 1.25 & 1.45 & 1.36 & 0.24 & 0.21 & 0.20 & 25 & 11 & 11 \\
Staged deployment & 1.32 & 1.45 & 1.27 & 0.22 & 0.16 & 0.19 & 25 & 11 & 11 \\
Tracking model weights & 1.35 & 1.43 & 1.33 & 0.17 & 0.30 & 0.49 & 25 & 11 & 11 \\
Internal audit & 1.12 & 1.73 & 1.36 & 0.23 & 0.14 & 0.20 & 25 & 11 & 11 \\
No unsafe open-sourcing & 1.24 & 1.36 & 1.27 & 0.25 & 0.28 & 0.27 & 25 & 11 & 11 \\
Researcher model access & 1.38 & 1.00 & 0.82 & 0.19 & 0.30 & 0.23 & 25 & 11 & 11 \\
API access to powerful models & 1.08 & 1.44 & 1.50 & 0.25 & 0.24 & 0.17 & 25 & 11 & 11 \\
Avoiding hype & 1.24 & 0.82 & 1.36 & 0.14 & 0.26 & 0.15 & 25 & 11 & 11 \\
Gradual scaling & 1.13 & 1.36 & 1.09 & 0.25 & 0.15 & 0.21 & 25 & 11 & 11 \\
Treat updates similarly to new models & 1.08 & 1.45 & 1.00 & 0.24 & 0.25 & 0.23 & 25 & 11 & 11 \\
Pre-registration of large training runs & 0.87 & 1.64 & 1.00 & 0.26 & 0.20 & 0.27 & 25 & 11 & 11 \\
Enterprise risk management & 0.71 & 1.38 & 1.11 & 0.33 & 0.26 & 0.20 & 25 & 11 & 11 \\
Treat internal deployments similar to external deployments & 1.10 & 1.17 & 0.50 & 0.23 & 0.17 & 0.34 & 25 & 11 & 11 \\
Notify a state actor before deployment & 0.55 & 1.45 & 1.00 & 0.25 & 0.21 & 0.27 & 25 & 11 & 11 \\
Notify affected parties & 0.80 & 0.80 & 1.33 & 0.26 & 0.58 & 0.33 & 25 & 11 & 11 \\
Inter-lab scrutiny & 1.22 & 0.33 & 0.17 & 0.13 & 0.49 & 0.54 & 25 & 11 & 11 \\
Avoid capabilities jumps & 0.89 & 0.67 & 0.17 & 0.27 & 0.56 & 0.48 & 25 & 11 & 11 \\
Notify other labs & 0.72 & 0.80 & 0.00 & 0.19 & 0.37 & 0.37 & 25 & 11 & 11 \\
        \bottomrule
    \end{tabular}
\endgroup
\normalsize
\newpage

Table 5: \textbf{Statement Statistics: By sector (AGI labs, all other respondents)} | Mean, standard error and sample size (\textit{n}) for each of the fifty items divided by respondents’ sector of work. Here we separate out AGI lab respondents from all other respondents, to correspond with Figure \ref{fig:figure_7}. The items are ordered by mean agreement score across all respondents.

\begingroup
    \centering
    \fontsize{6}{8}\selectfont
    \hspace*{-0cm}
    \begin{tabular}{l*{6}{r}}
        \toprule
        & \multicolumn{2}{c}{\textbf{Mean}} & \multicolumn{2}{c}{\textbf{Standard error}} & \multicolumn{2}{c}{\textbf{\textit{n}}} \\
        \cmidrule(lr){2-3} \cmidrule(lr){4-5} \cmidrule(lr){6-7}
        \textbf{AGI safety and governance practice} & \textbf{AGI Lab} & \textbf{Everyone else} & \textbf{AGI Lab} & \textbf{Everyone else} & \textbf{AGI Lab} & \textbf{Everyone else} \\
        \midrule
        Pre-deployment risk assessment & 1.96 & 1.82 & 0.04 & 0.11 & 25 & 22 \\
Dangerous capabilities evaluations & 1.92 & 1.91 & 0.06 & 0.06 & 24 & 22 \\
Third-party model audits & 1.80 & 1.82 & 0.13 & 0.08 & 25 & 22 \\
Safety restrictions & 1.92 & 1.73 & 0.06 & 0.12 & 25 & 22 \\
Red teaming & 1.76 & 1.73 & 0.17 & 0.10 & 25 & 22 \\
Monitor systems and their uses & 1.72 & 1.82 & 0.14 & 0.08 & 25 & 22 \\
Alignment techniques & 1.72 & 1.73 & 0.18 & 0.10 & 25 & 22 \\
Security incident response plan & 1.77 & 1.86 & 0.09 & 0.10 & 22 & 14 \\
Post-deployment evaluations & 1.81 & 1.54 & 0.09 & 0.24 & 21 & 13 \\
Report safety incidents & 1.71 & 1.68 & 0.18 & 0.10 & 24 & 22 \\
Safety vs. capabilities & 1.68 & 1.73 & 0.13 & 0.10 & 25 & 22 \\
Internal review before publication & 1.68 & 1.64 & 0.17 & 0.10 & 25 & 22 \\
Pre-training risk assessment & 1.56 & 1.73 & 0.17 & 0.19 & 25 & 22 \\
Emergency response plan & 1.68 & 1.59 & 0.15 & 0.11 & 25 & 22 \\
Protection against espionage & 1.68 & 1.62 & 0.19 & 0.14 & 22 & 13 \\
Pausing training of dangerous models & 1.36 & 1.91 & 0.21 & 0.06 & 25 & 22 \\
Increasing levels of external scrutiny & 1.62 & 1.59 & 0.16 & 0.11 & 24 & 22 \\
Publish alignment strategy & 1.63 & 1.50 & 0.11 & 0.14 & 19 & 14 \\
Bug bounty programs & 1.62 & 1.36 & 0.15 & 0.12 & 24 & 22 \\
Industry sharing of security information & 1.55 & 1.46 & 0.11 & 0.22 & 22 & 13 \\
Security standards & 1.27 & 1.79 & 0.30 & 0.10 & 22 & 19 \\
Publish results of internal risk assessments & 1.65 & 1.17 & 0.11 & 0.34 & 20 & 12 \\
Dual control & 1.70 & 1.33 & 0.11 & 0.19 & 20 & 12 \\
Publish results of external scrutiny & 1.59 & 1.45 & 0.11 & 0.21 & 22 & 11 \\
Military-grade information security & 1.28 & 1.59 & 0.25 & 0.13 & 25 & 22 \\
Board risk committee & 1.61 & 1.08 & 0.12 & 0.45 & 18 & 12 \\
Chief risk officer & 1.53 & 1.25 & 0.16 & 0.33 & 19 & 12 \\
Statement about governance structure & 1.57 & 1.31 & 0.13 & 0.26 & 21 & 13 \\
Publish views about AGI risk & 1.39 & 1.32 & 0.22 & 0.12 & 23 & 22 \\
KYC screening & 1.43 & 1.38 & 0.15 & 0.18 & 21 & 13 \\
Third-party governance audits & 1.30 & 1.42 & 0.22 & 0.19 & 20 & 12 \\
Background checks & 1.62 & 1.17 & 0.13 & 0.27 & 21 & 12 \\
Model containment & 1.25 & 1.41 & 0.24 & 0.14 & 24 & 22 \\
Staged deployment & 1.32 & 1.36 & 0.22 & 0.12 & 25 & 22 \\
Tracking model weights & 1.35 & 1.38 & 0.17 & 0.27 & 20 & 13 \\
Internal audit & 1.12 & 1.55 & 0.23 & 0.13 & 25 & 22 \\
No unsafe open-sourcing & 1.24 & 1.32 & 0.25 & 0.19 & 25 & 22 \\
Researcher model access & 1.38 & 0.90 & 0.19 & 0.18 & 24 & 21 \\
API access to powerful models & 1.08 & 1.47 & 0.25 & 0.14 & 24 & 19 \\
Avoiding hype & 1.24 & 1.09 & 0.14 & 0.16 & 25 & 22 \\
Gradual scaling & 1.13 & 1.23 & 0.25 & 0.13 & 23 & 22 \\
Treat updates similarly to new models & 1.08 & 1.23 & 0.24 & 0.17 & 24 & 22 \\
Pre-registration of large training runs & 0.87 & 1.32 & 0.26 & 0.18 & 23 & 22 \\
Enterprise risk management & 0.71 & 1.24 & 0.33 & 0.16 & 17 & 17 \\
Treat internal deployments similar to external deployments & 1.10 & 0.83 & 0.23 & 0.21 & 20 & 12 \\
Notify a state actor before deployment & 0.55 & 1.23 & 0.25 & 0.17 & 22 & 22 \\
Notify affected parties & 0.80 & 1.09 & 0.26 & 0.31 & 15 & 11 \\
Inter-lab scrutiny & 1.22 & 0.25 & 0.13 & 0.35 & 18 & 12 \\
Avoid capabilities jumps & 0.89 & 0.42 & 0.27 & 0.36 & 18 & 12 \\
Notify other labs & 0.72 & 0.36 & 0.19 & 0.28 & 18 & 11 \\
        \bottomrule
    \end{tabular}
\endgroup
\normalsize

\newpage
Table 6: \textbf{Statement Statistics: By gender} | Mean, standard error and sample size (\textit{n}) for each of the fifty items divided by respondents’ gender. These represent the two groups with sufficiently high sample sizes for analyses of group differences. The items are ordered by mean agreement score across all respondents.

\begingroup
    \centering
    \fontsize{6}{8}\selectfont
    \hspace*{1.5cm}
    \begin{tabular}{l*{6}{r}}
        \toprule
        & \multicolumn{2}{c}{\textbf{Mean}} & \multicolumn{2}{c}{\textbf{Standard error}} & \multicolumn{2}{c}{\textbf{\textit{\textit{n}}}} \\
        \cmidrule(lr){2-3} \cmidrule(lr){4-5} \cmidrule(lr){6-7}
        \textbf{AGI safety and governance practice} & \textbf{Men} & \textbf{Women} & \textbf{Men} & \textbf{Women} & \textbf{Men} & \textbf{Women} \\
        \midrule
        Pre-deployment risk assessment & 1.94 & 1.86 & 0.04 & 0.14 & 32 & 14 \\
Dangerous capabilities evaluations & 1.90 & 2.00 & 0.05 & 0.00 & 31 & 14 \\
Third-party model audits & 1.81 & 1.93 & 0.07 & 0.07 & 32 & 14 \\
Safety restrictions & 1.75 & 2.00 & 0.09 & 0.00 & 32 & 14 \\
Red teaming & 1.84 & 1.79 & 0.07 & 0.11 & 32 & 14 \\
Monitor systems and their uses & 1.62 & 2.00 & 0.12 & 0.00 & 32 & 14 \\
Alignment techniques & 1.62 & 2.00 & 0.15 & 0.00 & 32 & 14 \\
Security incident response plan & 1.75 & 1.92 & 0.09 & 0.08 & 24 & 12 \\
Post-deployment evaluations & 1.70 & 2.00 & 0.10 & 0.00 & 23 & 11 \\
Report safety incidents & 1.81 & 1.79 & 0.07 & 0.11 & 31 & 14 \\
Safety vs. capabilities & 1.59 & 1.86 & 0.11 & 0.10 & 32 & 14 \\
Internal review before publication & 1.59 & 1.86 & 0.14 & 0.10 & 32 & 14 \\
Pre-training risk assessment & 1.78 & 1.64 & 0.11 & 0.23 & 32 & 14 \\
Emergency response plan & 1.66 & 1.71 & 0.10 & 0.13 & 32 & 14 \\
Protection against espionage & 1.57 & 1.75 & 0.19 & 0.13 & 23 & 12 \\
Pausing training of dangerous models & 1.75 & 1.64 & 0.09 & 0.23 & 32 & 14 \\
Increasing levels of external scrutiny & 1.56 & 1.79 & 0.10 & 0.11 & 32 & 14 \\
Publish alignment strategy & 1.43 & 1.60 & 0.12 & 0.16 & 23 & 10 \\
Bug bounty programs & 1.53 & 1.57 & 0.10 & 0.14 & 32 & 14 \\
Industry sharing of security information & 1.35 & 1.73 & 0.13 & 0.14 & 23 & 11 \\
Security standards & 1.44 & 1.75 & 0.22 & 0.13 & 27 & 12 \\
Publish results of internal risk assessments & 1.33 & 1.82 & 0.14 & 0.12 & 21 & 11 \\
Dual control & 1.50 & 1.50 & 0.13 & 0.22 & 22 & 10 \\
Publish results of external scrutiny & 1.36 & 1.73 & 0.12 & 0.14 & 22 & 11 \\
Military-grade information security & 1.47 & 1.36 & 0.15 & 0.25 & 32 & 14 \\
Board risk committee & 1.40 & 1.60 & 0.22 & 0.16 & 20 & 10 \\
Chief risk officer & 1.58 & 1.25 & 0.16 & 0.18 & 19 & 12 \\
Statement about governance structure & 1.43 & 1.73 & 0.14 & 0.14 & 23 & 11 \\
Publish views about AGI risk & 1.33 & 1.50 & 0.12 & 0.17 & 30 & 14 \\
KYC screening & 1.23 & 1.50 & 0.16 & 0.15 & 22 & 12 \\
Third-party governance audits & 1.48 & 1.33 & 0.13 & 0.19 & 21 & 12 \\
Background checks & 1.42 & 1.22 & 0.16 & 0.28 & 24 & 9 \\
Model containment & 1.39 & 1.36 & 0.16 & 0.20 & 31 & 14 \\
Staged deployment & 1.34 & 1.57 & 0.15 & 0.17 & 32 & 14 \\
Tracking model weights & 1.27 & 1.33 & 0.18 & 0.26 & 22 & 12 \\
Internal audit & 1.28 & 1.43 & 0.16 & 0.17 & 32 & 14 \\
No unsafe open-sourcing & 1.38 & 1.07 & 0.18 & 0.32 & 32 & 14 \\
Researcher model access & 1.13 & 1.50 & 0.16 & 0.14 & 30 & 14 \\
API access to powerful models & 1.24 & 1.31 & 0.19 & 0.24 & 29 & 13 \\
Avoiding hype & 1.06 & 1.21 & 0.13 & 0.19 & 32 & 14 \\
Gradual scaling & 1.13 & 1.57 & 0.12 & 0.17 & 30 & 14 \\
Treat updates similarly to new models & 1.00 & 1.36 & 0.17 & 0.29 & 31 & 14 \\
Pre-registration of large training runs & 1.13 & 1.36 & 0.15 & 0.25 & 31 & 14 \\
Enterprise risk management & 0.90 & 1.23 & 0.24 & 0.20 & 21 & 13 \\
Treat internal deployments similar to external deployments & 0.95 & 1.30 & 0.17 & 0.26 & 22 & 10 \\
Notify a state actor before deployment & 0.93 & 1.14 & 0.18 & 0.21 & 30 & 14 \\
Notify affected parties & 1.07 & 1.10 & 0.18 & 0.31 & 15 & 10 \\
Inter-lab scrutiny & 0.72 & 0.82 & 0.27 & 0.23 & 18 & 11 \\
Avoid capabilities jumps & 0.62 & 1.00 & 0.20 & 0.44 & 21 & 9 \\
Notify other labs & 0.40 & 0.89 & 0.21 & 0.26 & 20 & 9 \\
        \bottomrule
    \end{tabular}
\endgroup
\normalsize

\newpage

\subsection*{Demographics}

Table 7: \textbf{Demographics of sample: Sector} | Percentage and frequency of respondents by sector. Note that respondents could report more than one sector

\begingroup
\begin{tabularx}{1\textwidth}{>{\raggedright\arraybackslash}X >{\raggedright\arraybackslash}X r r}
\toprule
\textbf{Sector} & \textbf{Sector subgroup} & \textbf{Percentage of total sample} & \textbf{Raw frequency} \\
\midrule
AGI lab & & 43.9\% & 25 \\
\midrule
Academia & & 22.8\% & 13 \\
\midrule
Civil society & & & \\
& Think tank & 10.5\% & 6 \\
& Nonprofit organization & 12.3\% & 7 \\
\midrule
Other & & & \\
& Other tech company & 1.8\% & 1 \\
& Government & 0\% & 0 \\
& Consulting firm & 1.8\% & 1 \\
& Other & 1.8\% & 1 \\
\midrule
Prefer not to say & & 5.3\% & 3 \\
\bottomrule
\end{tabularx}
\endgroup
\normalsize
\hfill
\newline

\begingroup
\centering
Table 8: \textbf{Demographics of sample: Gender} | Percentage and frequency of respondents by gender

\begin{tabular}{lll}
\toprule
{} & \textbf{Raw frequency} & \textbf{Percentage of total sample} \\
\textbf{Gender}            &       &            \\
\midrule
Man               &    32 &      62.7\% \\
Woman             &    14 &      27.5\% \\
Prefer not to say &     5 &       9.8\% \\
Another gender    &     0 &       0.0\% \\
\bottomrule
\end{tabular}
\par
\endgroup


\newpage
\section{Additional analyses}
\label{section:F}

\subsection*{Deviations from the pre-registered pre-analysis plan}

We pre-registered the survey on OSF (\url{https://osf.io/s7vhr}). We generally followed the pre-analysis plan. We present several additional top-line statistics that were not noted in the pre-analysis plan, such as how many statements received a majority of agreement responses. We did not conduct the pre-registered regression analyses to test for the effect of sector or gender due to the small sample size. We ran the pre-registered Mann-Whitney U and Chi-squared tests instead, with appropriate correction for multiple comparisons where applicable (using the Holm-Bonferroni correction). We did not run the Kolgmorov-Smirnov tests, since the Mann-Whitney U-test was more appropriate for the observed distributions.

\subsection*{Cluster analysis}

In an attempt to discover groups of response patterns within the population, we attempted to cluster respondents using their pattern of responses across questions and their reported demographic data. In line with our pre-analysis plan, we conducted k-means clustering on the dataset of responses and demographic labels (for the variables “gender” and “sector”). The aim of this analysis is to discover high-dimensional clusters or groups of response patterns within the population of respondents, and to visualize these in a more interpretable, low-dimensional manner. To achieve this, we performed a number of standard data pre-processing steps for dimensionality reduction techniques \cite{likas:2003}. 

We firstly pre-processed the data to remove respondents with missing demographic data. The gender and sector demographic variables were then transformed into binary features with one-hot encoding. In the final data pre-processing step, we standardized the data to ensure that the variables were approximately equally scaled (this was done using the \texttt{StandardScaler} functionality from the library \texttt{sklearn}). To partition the processed data for visualization, we employed the standard k-means clustering algorithm. In this algorithm, the number of clusters is a hyperparameter, which must be estimated or inferred. To select the optimal number of clusters in a principled manner, we employed two accepted methods – the Elbow method and silhouette analysis \cite{saputra:2020}– which evaluated the inertia and silhouette score of the model for a range of clusters \(n \in \{2,3,4,5,6,7,8,9,10\}\), where n represents the number of clusters). 

Based on this analysis, we found the optimal number of clusters to be four, and performed k-means clustering with four clusters accordingly. To visualize this clustered data, we first reduced the dimensionality of the embedded data to two dimensions (that is, two axes for visualization) using principal component analysis (PCA), and then visualized the results using a scatter plot.  We found the clusters to be poorly separated, implying that it is difficult to represent groups in this dataset in a low-dimensional manner (in support of this, the Elbow error metric was relatively high for all given numbers of clusters \(n \in \{2,3,4,5,6,7,8,9,10\}\). This could be due to a number of reasons: the relatively small sample of this population, poor scaling of the variables of the data (as discussed above), or the presence of non-convex clusters.

All of the code for this analysis, along with some instructive visualizations, can be found on OSF (\url{https://osf.io/s7vhr}).

\newpage

\bibliographystyle{abbrv}
\small \bibliography{ms}

\end{document}